\documentclass[aps,amsmath,amssymb,superscriptaddress,prb,10pt,twocolumn,eqrefnum]{revtex4-2}

\usepackage{graphicx}
\usepackage{dcolumn}
\usepackage{bm}
\renewcommand{\vec}[1]{\mathbf{#1}}
\usepackage{physics}
\usepackage{mathrsfs,amsthm}
\usepackage{xcolor}
\usepackage{esvect}
\usepackage{appendix}
\usepackage[normalem]{ulem}
\usepackage{hyperref}

\DeclareMathAlphabet{\mathpzc}{OT1}{pzc}{m}{it}
\begin{document}

\preprint{APS/123-QED}

\title{Spin-resolved Josephson diode effect through strongly spin-polarized conical magnets}

\author{Danilo Nikoli\'c}
\email{danilo.nikolic@uni-greifswald.de}
\affiliation{Institut f\"ur Physik, Universit\"at Greifswald, Felix-Hausdorff-Straße 6, 17489 Greifswald, Germany}
\author{Niklas L. Schulz}%
 \email{niklas.schulz@uni-greifswald.de}
\affiliation{Institut f\"ur Physik, Universit\"at Greifswald, Felix-Hausdorff-Straße 6, 17489 Greifswald, Germany}
\author{Alexander I. Buzdin}
\affiliation{University Bordeaux, LOMA UMR-CNRS 5798, F-33405 Talence Cedex, France}
\affiliation{Institute for Computer Science and Mathematical Modeling, World-Class Research Center Digital Biodesign and Personalized Healthcare, Sechenov First Moscow State Medical University, Moscow 119991, Russia}
\author{Matthias Eschrig}%
 \email{matthias.eschrig@uni-greifswald.de}
\affiliation{Institut f\"ur Physik, Universit\"at Greifswald, Felix-Hausdorff-Straße 6, 17489 Greifswald, Germany}

\date{\today}

\begin{abstract}
We present a theoretical study of the spin-resolved Josephson diode effect in junctions comprising strongly spin-polarized conical magnets (FM) coupled to singlet superconductors (SC). The system is treated by making use of the Gor'kov and quasiclassical Green's function methods. Modeling the SC/FM interfaces as {spin-dependent} $\delta$-potentials, we apply our model to an SC/FM/SC junction and account for the Josephson current-phase relation (CPR). The nontrivial coupling between the spin bands in the conical magnet gives rise to a strong Josephson diode effect with an efficiency {greater} than $40\%$. The effect essentially depends on the quantum spin-geometric phase that enters the Josephson CPR in a very similar manner to the superconducting phase difference. The former is generated non-locally by the intrinsically noncoplanar spin arrangement of the conical {magnet}, which breaks the time-reversal and inversion symmetries. Strong spin polarization and a helical pitch of the conical magnet comparable to the superconducting coherence length are essential for the effect. We perform a harmonic analysis of the Josephson CPR and interpret the effect in terms of coherent transfer of multiple equal-spin triplet Cooper pairs across the conical magnet.
\end{abstract}

\maketitle

\section{Introduction}\label{sec:Intro}
The creation and control of long-range equal-spin triplet currents are of essential importance both for fundamental understanding and for application in superconducting spintronics~\cite{eschrigSingletTripletMixingSuperconductor2004,buzdinProximityEffectsSuperconductorferromagnet2005,bergeretOddTripletSuperconductivity2005,eschrigSpinpolarizedSupercurrentsSpintronics2011,eschrigSpinpolarizedSupercurrentsSpintronics2015,linderSuperconductingSpintronics2015,birgereview2018,linderOddfrequencySuperconductivity2019,yangBoostingSpintronicsSuperconductivity2021,caiSuperconductorFerromagnetHeterostructures2023}. Typical examples of the systems featuring such correlations are superconductor (SC)-ferromagnet (FM) hybrids among which strongly spin-polarized itinerant ferromagnets sandwiched between two spin-singlet superconductors are {of particular interest}. Due to the strong spin polarization in such devices, the phase-coherence {over mesoscopic distances} is maintained only within each spin band, however, not between them~ \cite{eschrigTheoryHalfMetalSuperconductor2003,eschrigTripletSupercurrentsClean2008,greinSpinDependentCooperPair2009,eschrigScatteringProblemNonequilibrium2009,greinInverseProximityEffect2013,eschrigSpinpolarizedSupercurrentsSpintronics2015,houzetQuasiclassicalTheoryDisordered2015,bobkovaGaugeTheoryLongrange2017,ouassouTripletCooperPairs2017,eschrigTheoryAndreevBound2018}. As it is known, SC/FM systems with a conventional superconducting order parameter can host pair amplitudes, which are classified into singlet and triplet spin correlations {\cite{tokuyasuProximityEffectFerromagnetic1988,demlerSuperconductingProximityEffects1997,bergeretLongRangeProximityEffects2001,eschrigSymmetriesPairingCorrelations2007,tanakaTheoryProximityEffect2007}}. Due to the proximity effect, spin-singlet correlations from the superconductor can penetrate the ferromagnet, converting into triplet pair correlations, which are now inequivalent and can be classified into short-range and long-range correlations~\cite{bergeretLongRangeProximityEffects2001}. Considering the magnetization in the ferromagnet as a spin quantization axis allows identifying the equal-spin $\uparrow\uparrow$- and $\downarrow\downarrow$-pairs as long-ranged within a respective spin band, and the third, mixed-spin, triplet as short-ranged, just like the singlet amplitude. The short-range nature of the latter two is due to the decoherence induced by the exchange splitting between the bands.

The creation and control of equal-spin correlations is allowed due to two fundamental processes occurring in the vicinity of an SC/FM interface~\cite{eschrigSpinpolarizedSupercurrentsSpintronics2011}. First, the spin-mixing (or spin-dependent-phase-shift) effect due to the spin polarization of the interface converts spin-singlet pairs from the SC into mixed-spin triplet correlations in the FM. Second, if a noncollinear spin arrangement is present in the FM, the spin-rotation mechanism turns the short-range mixed-spin pairs into the long-range equal-spin pairs. These two mechanisms of production and control of spin-triplet supercurrents have been proven  experimentally~\cite{keizerSpinTripletSupercurrent2006,khaireObservationSpinTripletSuperconductivity2010,anwarLongrangeSupercurrentsHalfmetallic2010,robinsonControlledInjectionSpinTriplet2010,Glick2018,Caruso2019,Aguilar2020}.

According to the mechanism outlined above, a noncollinear magnetization profile gives rise to long-range triplet supercurrents that are essential for superconducting spintronics. However, new functionalities appear when the spin texture is not only noncollinear,  but in fact noncoplanar. Specifically, such a spin profile, combined with a strong spin polarization of the material, leads to an effective decoupling of the Josephson phases in the two spin bands. This decoupling, in turn, opens a new channel of control through geometric phases~\cite{eschrigSymmetriesPairingCorrelations2007,eschrigTripletSupercurrentsClean2008}, which are coupled to the Josephson phases with opposite signs in the two spin bands~\cite{greinSpinDependentCooperPair2009}. Previous work, Refs.~\cite{schulz2025_prl,schulz2025_prb}, has reported on the essential role of nonlocal geometric phases, determined by a noncoplanar magnetization profile in {strongly spin-polarized} ferromagnetic trilayers, in the Josephson diode effect. In this article, we further exploit this important concept, proposing an alternative platform based on intrinsically noncoplanar magnetic materials, such as conical magnets. {Similar systems have been discussed in the context of helical superconductivity~\cite{bulaevskiiHelicalOrderingSpins1980,mengNonuniformSuperconductivityJosephson2019}, spin-triplet generation~\cite{bergeretLongRangeProximityEffects2001,volkovNeel2005,sosninSuperconductingPhaseCoherent2006a,majeedModellingTripletProximity2025,Bovkova2025}, Josephson effect~\cite{volkovOddTripletSuperconductivity2006,fominovJosephsonEffectDue2007,champelCycloidalSpiral2008,wittJosephsonJunctionsIncorporating2012}, or spin-valve effects~\cite{champelEffectInhomogeneousExchange2005,spuriSignatureLongrangedSpin2024}, just to mention a few. However, most of these works assumed weak spin polarization of the ferromagnet, which reflects itself in the equal normal density of states of the two spin bands. In contrast, in this article, we emphasize the crucial role of the different densities of states in the anomalous and the Josephson diode effect.}

If the Josephson junction is invariant under the time reversal and inversion, the current-phase relation (CPR) of the junction exhibits an odd symmetry with respect to reversal of the superconducting phase difference, $I(-\Delta\chi)=-I(\Delta\chi)\implies I(\Delta\chi=0)=0$, and this effect is known as the normal Josephson effect.~\cite{golubovCurrentphaseRelationJosephson2004}. However, if time reversal and inversion symmetries are broken, the anomalous Josephson effect may appear, $I(-\Delta\chi)\neq I(\Delta\chi) \implies I(\Delta\chi=0)\neq
0$~\cite{GeshkenbeinLarkin1986,Yip1995,Sigrist1998,Buzdin2008}. Furthermore, if higher harmonics in the CPR are present, the critical current in one direction $(+)$ may differ from that in the opposite direction $(-)$~\cite{nadeemSuperconductingDiodeEffect2023}. This effect is known as the Josephson diode effect (JDE) and recently it has attracted considerable attention as the subject of intense experimental~\cite{andoObservationSuperconductingDiode2020, baumgartnerSupercurrentRectificationMagnetochiral2022, costaSignReversalJosephson2023, gutfreundDirectObservationSuperconducting2023, houUbiquitousSuperconductingDiode2023, nadeemSuperconductingDiodeEffect2023, strambiniSuperconductingSpintronicTunnel2022, trahmsDiodeEffectJosephson2023,reinhardtLinkSupercurrentDiode2024}  and theoretical investigations~\cite{greinSpinDependentCooperPair2009,costaSignReversalJosephson2023, daidoIntrinsicSuperconductingDiode2022, fominovAsymmetricHigherharmonicSQUID2022, haltermanSupercurrentDiodeEffect2022, hePhenomenologicalTheorySuperconductor2022, ilicTheorySupercurrentDiode2022, karabassovHybridHelicalState2022, kopasovGeometryControlledSuperconducting2021, misakiTheoryNonreciprocalJosephson2021, tanakaTheoryGiantDiode2022, yuanSupercurrentDiodeEffect2022, zhangGeneralTheoryJosephson2022, zinklSymmetryConditionsSuperconducting2022,soutoJosephsonDiodeEffect2022,steinerDiodeEffectsCurrentBiased2023,costaMicroscopicStudyJosephson2023,kopasovAdiabaticPhasePumping2023,Meyer2024,putilovNonreciprocalElectronTransport2024,tjernshaugenSuperconductingPhaseDiagram2024,patil2024,schulz2025_prl,schulz2025_prb}. It is customarily characterized by the so-called diode efficiency defined as  $ \eta_\mathrm{e}=(I^+-|I^-|)/(I^+ + |I^-|)$, where $I^\pm$ refers to the critical current in the corresponding direction.

In this article, we discuss the appearance of the Josephson diode effect in junctions involving strongly spin-polarized magnetic materials featuring inhomogeneous noncoplanar spin textures. We first consider the general theory in the so-called adiabatic regime discussed below, then we apply it to a conical magnet placed between two BCS superconductors. The anomalous and the Josephson diode effects have already been studied in similar systems utilizing the Bogoliubov-de Gennes formalism~\cite{mengNonuniformSuperconductivityJosephson2019,Andersen2024,kamraNonreciprocalJosephsonCurrent2024}. The quasiclassical Green's function treatment of such a system was reported in Ref.~\cite{bobkovaGaugeTheoryLongrange2017}; however, due to the absence of higher harmonics in CPR only the anomalous Josephson effect ("$\varphi_0$-junction") was observed. In this paper, we apply the appropriate boundary conditions, which allow us to account for higher harmonics in the CPR, yielding the significant JDE with an efficiency of up to $\sim 41\%$.

\begin{figure}[t!]
    \centering
    \includegraphics[width=0.9\linewidth]{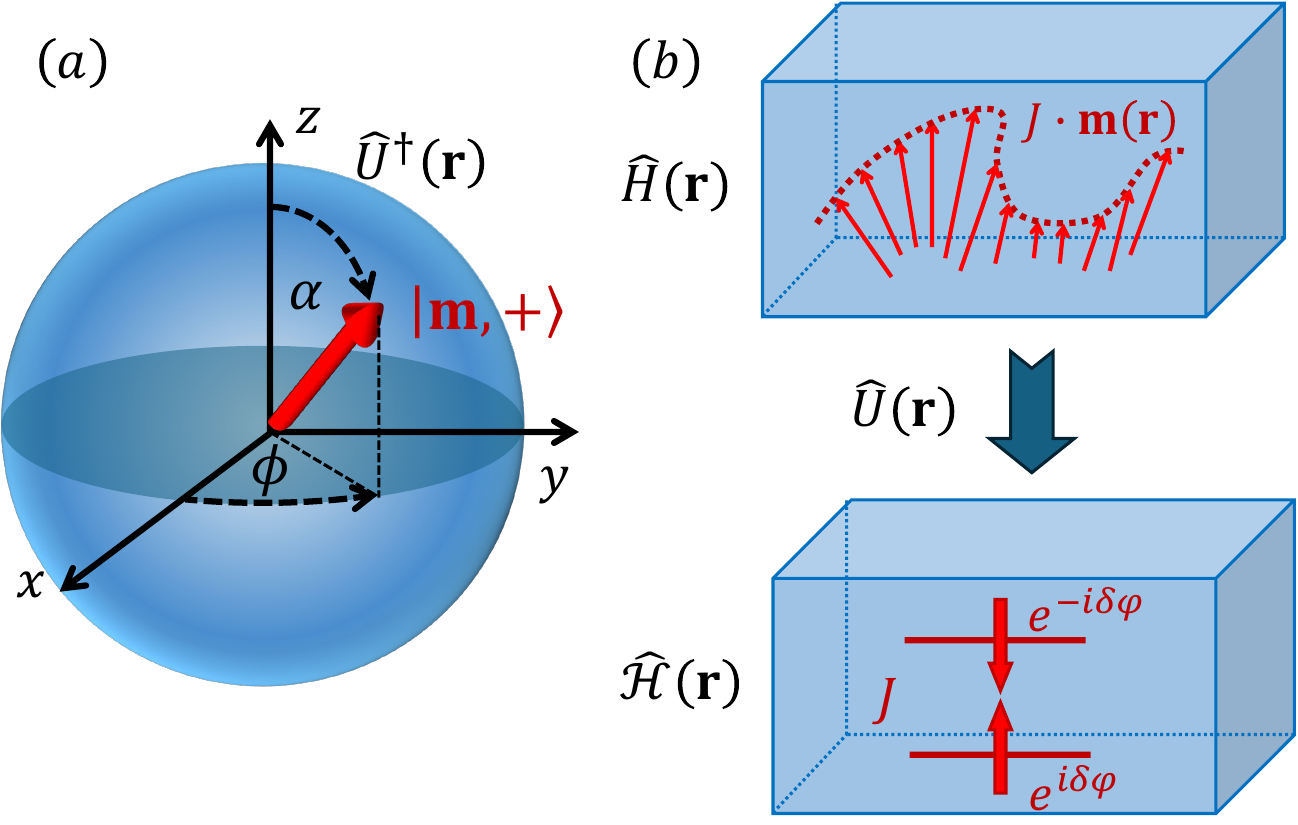}
    \caption{Panel (a): A representation of $\ket{\vec{m},+}$ state [see Eq.~\eqref{eqn:m_state}] on the Bloch sphere, where the north and the south pole denote the $\ket{\uparrow}$ and  $\ket{\downarrow}$ spin states, respectively. Panel (b): An illustration of the adiabatic approximation used throughout: an inhomogeneous ferromagnet with slowly varying magnetization can be locally represented as a homogeneous state with different U(1) gauge phases acquired by the spin-$\uparrow$ and spin-$\downarrow$ states.}
    \label{fig:Spin_gauge_field}
\end{figure}

To qualitatively illustrate the adiabatic approximation that is crucial for the JDE in inhomogeneous magnets, let us consider a single electron interacting with a localized spin via exchange interaction
\begin{equation}
\label{eqn:F_Hamiltonian}
\hat{H} = \frac{\vec{\hat{p}}^2}{2m} - J\vec{m}(\vec{r})\cdot\bm{\hat{\sigma}},
\end{equation}
where $\vec{m}(\vec{r})=[\sin\alpha(\vec{r})\cos\phi(\vec{r}),\sin\alpha(\vec{r})\sin\phi(\vec{r}),\cos\alpha(\vec{r})]$ is the spin texture of the material. The spin eigenstates of the Hamiltonian above are 
\begin{equation}\label{eqn:m_state}
\begin{split}
&\ket{\vec{m},+}=\cos(\frac{\alpha}{2})\ket{\uparrow}+\sin(\frac{\alpha}{2})e^{i\phi}\ket{\downarrow},\\
&\ket{\vec{m},-}=-\sin(\frac{\alpha}{2})e^{-i\phi}\ket{\uparrow}+\cos(\frac{\alpha}{2})\ket{\downarrow}.
\end{split}
\end{equation}
 They can be obtained as spin rotations on the Bloch sphere [see Fig.~\ref{fig:Spin_gauge_field}(a)], $\ket{\vec{m},\pm}=\hat{U}^\dagger(\vec{r})\ket{\uparrow\!/\!\downarrow}$, where
\begin{equation}
\label{eqn:U}
    \hat{U}(\vec{r}) = \exp[-i\frac{\alpha(\vec{r})}{2}\vec{n}(\vec{r})\cdot\bm{\hat{\sigma}}],
\end{equation}
with $\vec{n}=(\vec{m}\times \vec{e}_3)/|\vec{m}\times \vec{e}_3|=\sin{\left[\phi(\vec{r})\right]}\vec{e}_1-\cos{\left[\phi(\vec{r})\right]}\vec{e}_2$. {The overlap between two states on the Bloch sphere $\braket{\vec{m'},\pm}{\vec{m},\pm}=\mel{\uparrow\!/\!\downarrow}{\hat{U}(\vec{r'})\hat{U}^\dagger(\vec{r})}{\uparrow\!/\!\downarrow}$ can be evaluated, using Eq.~\eqref{eqn:U}, explicitly as
\begin{equation}
    \braket{\vec{m'},\pm}{\vec{m},\pm} = \cos\frac{\alpha}{2}\cos\frac{\alpha^\prime}{2} + \sin\frac{\alpha}{2}\sin\frac{\alpha^\prime}{2}e^{\pm i(\phi-\phi^\prime)},
\end{equation}
where we use the abbreviation $\alpha\equiv\alpha(\vec{r})$ and $\alpha^\prime\equiv\alpha(\vec{r}^\prime)$, and analogously for $\phi$. Assuming slow spatial variations of the spin texture (adiabatic condition), we obtain 
\begin{equation}
\bra{\vec{m}',\pm}\ket{\vec{m},\pm}\approx 1 \pm i\delta\varphi\approx e^{\pm i\delta\varphi},
\end{equation}
 where 
 \begin{equation}
 \delta\varphi=\sin^2\left(\frac{\alpha}{2}\right)[\phi(\vec{r})-\phi(\vec{r}^\prime)]\approx \sin^2\left(\frac{\alpha}{2}\right)(\delta\vec{r}\cdot\bm{\nabla}\phi)
 \end{equation}
 is the so-called \textit{(adiabatic) spin gauge phase} with $\delta\vec{r}=\vec{r}-\vec{r}^\prime$ being the relative coordinate.} For a conical magnet, where $\alpha=$const., the adiabatic spin gauge phase can be integrated between two spatial points $\vec{r}_1$ and $\vec{r}_2$, leading to a {\it spin geometric phase} 
 \begin{equation}
 \Delta \varphi_{s}=\sin^2\left(\frac{\alpha}{2}\right) \left[\phi(\vec{r}_2)-\phi(\vec{r}_1)\right].
 \end{equation}
 {We make a few observations here. First, it is important to stress that in mesoscopic superconductors, due to the nonlocality of Cooper pairs that extend over the superconducting coherence length, the spin geometric phase $\Delta\varphi_s$ introduced above is \textit{nonlocal} on atomic length scales. Second, we note that in strongly spin-polarized materials, the equal-spin pairing correlations of the two spin bands acquire the opposite spin geometric phases. As we show in Sec.~\ref{sec:Results}, the difference between these phases enters the Josephson CPR very similarly to the superconducting phase difference, proving to be crucial for the Josephson diode effect and confirming the mechanism introduced in Refs.~\cite{schulz2025_prl,schulz2025_prb}. Finally, integrating the spin gauge phase $\delta\phi$ from $\varphi(\vec{r}_1)=0$ to $\varphi(\vec{r}_2)=2\pi$ yields the spin geometric phase of $\Delta\varphi_s=2\pi\sin^2(\alpha/2)$. This value corresponds to one half of the solid angle of the corresponding cone, namely, $\Delta\varphi_s=\Omega/2$, where $\Omega=4\pi\sin^2(\alpha/2)$, leading to a geometric interpretation of the quantity $\Delta\varphi_s$.} A schematic view of the adiabatic approximation is shown in Fig.~\ref{fig:Spin_gauge_field}(b).

 The paper is organized as follows. In Sec.~\ref{sec:Theory}, we present the theoretical framework suited for strongly spin-polarized inhomogeneous spin textures, which is based on the Gor'kov and quasiclassical Green's function techniques. In Sec.~\ref{sec:System}, we describe the system under study and apply the previously developed formalism to a conical magnet coupled to two superconductors via spin-polarized interfaces. In Sec.~\ref{sec:Results}, we present the results of our numerical calculations and detailed analysis of the Josephson current-phase relation. Finally, we give the concluding remarks in Sec.~\ref{sec:Conclusions}. 
 
\section{Quasiclassical theory}\label{sec:Theory}
\subsection{Gor'kov equations}
Mesoscopic superconductivity is successfully described by the many-body Green's function method. The cornerstone of this formalism is the so-called Gor'kov Green's function theory,  formulated in Gor'kov-Nambu (particle-hole) space. If one has to deal with an additional spin degree of freedom (e.g., due to spin-dependent interactions), this formalism invokes   the larger Gor'kov-Nambu $\otimes$ spin space. In this case, the equation of motion (EOM) for Gor'kov Green's function in equilibrium reads~\cite{Gorkov1958,eschrigSpinpolarizedSupercurrentsSpintronics2015}
\begin{equation}\label{eqn:Gorkov}
	\left[i\omega_n\check{\tau}_3\!-\!\xi(\vec{\hat{p}})+\vec{J}(\vec{r})\!\cdot\!\bm{\bar{\sigma}}-\check{\Delta}(\vec{r})\right]\!\check{\tau}_3\check{G}(\vec{r,r}')=\delta(\vec{r}\!-\!\vec{r}'),
\end{equation}
where $\omega_n=(2n+1)\pi k_BT$ are the Matsubara {energies} with $n=0,\pm 1,\pm 2, \dots$ and temperature $T$, $\xi(\vec{\hat{p}})=E(\vec{\hat{p}})-\mu$ with $\vec{\hat{p}}=-i\hbar\bm{\nabla}-e\vec{A}\check{\tau}_3$ denoting the {kinetic} momentum (including the vector potential $\vec{A}$) and $\mu$ the chemical potential. Here we do not restrict ourselves to the parabolic dispersion but we consider a general energy dispersion  relation $E(\vec{\hat{p}})$. Finally, $\vec{J}= J [\sin\alpha(\vec{r})\cos\phi(\vec{r}),\sin\alpha(\vec{r})\sin\phi(\vec{r}),\cos\alpha(\vec{r})]$ is the exchange field characterizing the spin texture of the material. In addition, the BCS self-energy (gap matrix) $\check{\Delta}$ and the spin matrix $\bar{\bm{\sigma}}$ have the forms 
\begin{equation}
    \check{\Delta} = \left(
    \begin{array}{cccc}
        0 & \hat{\Delta} \\
        \tilde{\Delta} & 0
    \end{array}{}
    \right)
    \quad \text{and}\quad
    \bm{\bar{\sigma}} = \left(
    \begin{array}{cccc}
        \bm{\hat{\sigma}} & 0 \\
        0 & \bm{\hat{\sigma}}^\ast
    \end{array}{}
    \right).
\end{equation}
We consider spin-singlet BCS superconductors for which $\tilde{\Delta}=\hat{\Delta}^\ast$ and $\hat{\Delta}=\Delta i\hat{\sigma}_2$. In the expressions above $\hat{\tau}_i$ and $\hat{\sigma}_i$ are the Pauli matrices in Gor'kov-Nambu and spin space, respectively, and check symbol ($\check{~}$) denotes a matrix in combined space, i.e., $\check{\tau}_i=\hat{\tau}_i\otimes\hat{\sigma}_0$ and $\check{\sigma}_i=\hat{\tau}_0\otimes\hat{\sigma}_i$. Finally, the Green's function is built from basis $\check{\Psi}=(\hat{\psi}_\uparrow,\hat{\psi}_\downarrow,\hat{\psi}^\dagger_\uparrow,\hat{\psi}^\dagger_\downarrow)$ having a $4\times 4$ matrix structure in combined space
\begin{equation}\label{eqn:GF}
    \check{G} = \left(
    \begin{array}{cccc}
         \hat{G} & \hat{F}\\
         \hat{\tilde{F}}& \hat{\tilde{G}}
    \end{array}{}
    \right)
    \quad\text{with}\quad
    \hat{F} = \left(
    \begin{array}{cccc}
         F_{\uparrow\uparrow} & F_{\uparrow\downarrow}\\
         F_{\downarrow\uparrow}& F_{\downarrow\downarrow}
    \end{array}{}
    \right).
\end{equation}

Our main objective in this section is to formulate the quasiclassical theory applicable to mesoscopic hybrid structures involving strongly spin-polarized magnetic materials (FM) for which the exchange field is comparable to the Fermi energy, $J\sim E_F$. For this purpose, we follow Refs.~\cite{greinSpinDependentCooperPair2009,greinTheorySuperconductorferromagnetPointcontact2010,bobkovaGaugeTheoryLongrange2017}. The developed theory relies on the assumptions that (i) FM is strongly spin-polarized and (ii) the spatial variations of the spin texture are slow, i.e., $\vec{J}(\vec{r})$ is nonlocal on the atomic length scales comparable to the Fermi wavelength $\lambda_F$. As we will show below, these assumptions allow for neglecting mixed-spin correlations due to the decoherence between the two spin {bands over mesoscopic distances}, keeping only the equal-spin correlations, $\uparrow\uparrow$ and $\downarrow\downarrow$.

Henceforth, we focus on the EOM in the FM layer and set $\Delta\equiv 0$ since there is no BCS coupling. To derive the effective equations, as a first step, we locally pass to the frame of reference in which $\vec{J}(\vec{r})$ is aligned with the $z$ axis, by performing the following SU(2) gauge transformation:
\begin{equation}
     \check{U}(\vec{r})(\vec{J}\cdot\bm{\bar{\sigma}}) \check{U}^\dagger(\vec{r}) = J\check{\sigma}_3.
\end{equation}
The introduced spin rotation operator $\check{U}$ has a $4 \times 4$ matrix structure in Gor'kov-Nambu $\otimes$ spin space:
\begin{equation}
    \check{U} = \left(
    \begin{array}{cc}
        \hat{U} & 0 \\
        0 & \hat{U}^\ast
    \end{array}{}\right)\quad\text{with}\quad \hat{U}=e^{-i\alpha(\vec{r})\vec{n}\cdot\bm{\hat{\sigma}}/2},
\end{equation}
where $\alpha$ is the polar angle and $\vec{n}$ is the rotation axis. The latter depends on the spin texture of the material $\vec{J}(\vec{r})$ as
\begin{equation}
    \vec{n} = \frac{\vec{J}\times\vec{e}_3}{|\vec{J}\times\vec{e}_3|}=\sin{\phi(\vec{r})}\vec{e}_1-\cos{\phi(\vec{r})}\vec{e}_2.
\end{equation}
Accordingly, the basis transforms as $\bar{\Psi}=\check{U}\check{\Psi}$ and the Gor'kov equations~\eqref{eqn:Gorkov} in the FM layer adopt the form
\begin{equation}\label{eqn:Gorkov_L}
	\bigg(\!i\omega_n\check{\tau}_3-\xi(\vec{\hat{p}})+i\frac{\check{A}_j\hat{v}_j}{2}+J\check{\sigma}_3\!\bigg)\check{\tau}_3\bar{G}(\vec{r,r}')\!=\!\delta(\vec{r}\!-\!\vec{r}'),
\end{equation}
where $\hbar\vec{\hat{v}}=\frac{\partial \xi}{\partial\vec{\hat{p}}}$ and a new SU(2) field is introduced, $\check{A}_i=\hbar\check{U}\nabla_i\check{U}^\dagger$. Here, we apply the adiabatic approximation for the first time. Namely, assuming that the $\vec{J}(\vec{r})$ is slowly varying {over mesoscopic length scales}, we only keep terms linear in the gradients of $\check{U}$. As shown below, the adiabatic condition will be employed several times, and in the subsequent subsection, we will quantify it by considering relevant length scales.

Equation \eqref{eqn:Gorkov_L} is the so-called left Gor'kov-Dyson equation. Analogously, we can write down the right Gor'kov-Dyson equation, namely the one acting on the second argument of the Gor'kov Green's function, $\vec{r}'$. Repeating the same procedure as above, we can introduce an analogous gauge transformation, arriving at the following equation:
\begin{equation}\label{eqn:Gorkov_R}
    \check{\tau}_3\bar{G}(\vec{r,r}')\!\bigg(\!i\omega_n\check{\tau}_3-\xi(\vec{\hat{p}'})-i\frac{\check{A}'_j\hat{v}_j}{2}+J\check{\sigma}_3\!\bigg)\!=\!\delta(\vec{r}\!-\!\vec{r}'),
\end{equation}
with the same notation as before. Note that here all operators act on the left, e.g., $\vec{\hat{p}'}=-i\hbar\overleftarrow{\bm{\nabla}}'-e\vec{A}\hat{\tau}_3$.

\subsection{Eilenberger equation}
The Gor'kov formalism presented above is fully microscopic, however not very practical for treating inhomogeneous mesoscopic problems. Instead, we formulate a quasiclassical theory which is typically more suitable in such situations~\cite{larkinQuasiclassicalMethodTheory1969,Eilenberger1968,belzigQuasiclassicalGreensFunction1999}. {Prior} to the quasiclassical approximation, we once more apply the adiabatic condition by keeping only the adiabatic gauge field which enters Eqs.~\eqref{eqn:Gorkov_L} and \eqref{eqn:Gorkov_R}~\cite{Bliokh2005,fujitaGaugeFieldsSpintronics2011,tatara2019}
\begin{equation}\label{eqn:addiabatic}
    \check{A}_j \hat{v}_j\!\approx\!\frac{\hbar}{2}(\hat{\tau}_3\otimes\hat{\sigma}_3)M_{3j}v_j~\text{with}~ M_{3j}\!=\!\Tr[\hat{\sigma}_3\hat{U}\nabla_j\hat{U}^\dagger]. 
\end{equation}

To derive the EOM for the quasiclassical propagator, we first parameterize the Gor'kov Green's function [see Eq.~\eqref{eqn:GF}] in a standard manner~\cite{sereneQuasiclassicalApproachSuperfluid1983,eschrigSpinpolarizedSupercurrentsSpintronics2015}
\begin{equation}
    \bar{G}=\left(
    \begin{array}{cc}
        G_0\hat{\sigma}_0+\bm{G}\cdot\bm{\hat{\sigma}} &  (F_0+\bm{F}\cdot\bm{\hat{\sigma}})i\hat{\sigma}_2\\
     (\tilde{F}_0+\bm{\tilde{F}}\cdot\bm{\hat{\sigma}}^\ast)i\hat{\sigma}_2    & \tilde{G}_0\hat{\sigma}_0+{\bm{\tilde{G}}\cdot\bm{\hat{\sigma}^\ast}} 
    \end{array}{} \right),
\end{equation}
and identify the mixed- and the equal-spin contributions to it, respectively,
\begin{equation}\label{eqn:G_MS}
\bar{G}_{\rm MS}\!=\!\left(
    \begin{array}{cc}
        G_1\hat{\sigma}_1+G_2\hat{\sigma}_2 & (F_0\hat{\sigma}_0+F_3\hat{\sigma}_3)i\hat{\sigma}_2 \\
       (\tilde{F}_0\hat{\sigma}_0+\tilde{F}_3\hat{\sigma}_3) i\hat{\sigma}_2  & \tilde{G}_1\hat{\sigma}_1-\tilde{G}_2\hat{\sigma}_2
    \end{array}\right),
\end{equation}
\begin{equation}\label{eqn:G_ES}
\bar{G}_{\rm ES}\!=\!\left(
    \begin{array}{cc}
        G_0\hat{\sigma}_0+G_3\hat{\sigma}_3 & (F_1\hat{\sigma}_1+F_2\hat{\sigma}_2)i\hat{\sigma}_2 \\
       (\tilde{F}_1\hat{\sigma}_1-\tilde{F}_2\hat{\sigma}_2) i\hat{\sigma}_2  & \tilde{G}_0\hat{\sigma}_0+\tilde{G}_3\hat{\sigma}_3
    \end{array}\right).
\end{equation}
As a second step, we employ the Wigner (mixed) representation, where the Green's function's spatial dependence is decomposed into the center-or-mass, $\vec{R}=(\vec{r}+\vec{r'})/2$, and relative coordinate, $\vec{x}=\vec{r-r}'$, contributions. {Subtracting Eqs.~\eqref{eqn:Gorkov_L} and \eqref{eqn:Gorkov_R} and applying a gradient expansion~\cite{sereneQuasiclassicalApproachSuperfluid1983} to linear order in gradients with respect to $\vec{R}$} yields the left-right-subtracted Gor'kov-Dyson equation:
\begin{equation}\label{eqn:Gorkov_LR}
    i\hbar\vec{v}\bm{\nabla}_\vec{R}\check{\tau}_3\bar{G}+\bigg[i\omega_n\check{\tau}_3+i\frac{\hbar}{4}(\hat{\tau}_3\otimes\hat{\sigma}_3)M_{3j}v_j+J\check{\sigma}_3,\check{\tau}_3\bar{G}\bigg]\!=\!0,
\end{equation}
with the same notation as before with $[\bullet\,,\bullet]$ denoting the commutator.

The equation above can be used to mathematically quantify the adiabatic condition used throughout. Inserting Eqs.~\eqref{eqn:G_MS} and \eqref{eqn:G_ES} into Eq.~\eqref{eqn:Gorkov_LR} and assuming 
that the main contribution to the Green's function comes from the excitations residing in the vicinity of the Fermi surface, it can be shown that the relevant length scales for the mixed-spin and equal-spin triplet amplitudes are $\xi_{\rm MS}\sim \hbar v_F/J$ and $\xi_{\rm ES}\sim {|M_{3j}|^{-1}}$, respectively. Therefore, the adiabatic condition is quantified by the relation $|\hbar v_FM_{3j}/J|\ll 1$ and it is fulfilled in the case of strongly spin-polarized, $J\sim E_F$, and/or slowly varying spin textures, {$|\lambda_F\partial_\vec{R}\hat{U}/\hat{U}|\ll 1$ in the FM region}. Here, $v_F$ denotes the Fermi velocity.

Now we proceed with the derivation of an effective quasiclassical theory. As pointed out above, the adiabatic approximation allows for neglecting mixed-spin correlations. Namely, in the case of a strongly spin-polarized ferromagnet, the phase coherence of pair correlations is maintained only within each spin band but not between them. In other words, the $\bar{G}_\mathrm{MS}$ contribution to Eq.~\eqref{eqn:Gorkov_LR} can be neglected, and the remaining, $\bar{G}_\mathrm{ES}$, part can be projected onto the respective spin band, arriving at 
\begin{equation}\label{eqn:Gorkov_ES}    i\hbar\vec{v}_\sigma\cdot\bm{\nabla}_\vec{R}\hat{\tau}_3\hat{\mathcal{G}}_\sigma+\bigg[i\omega_n\hat{\tau}_3+i\zeta_\sigma\frac{\hbar}{4}M_{3j}v_{\sigma j},\hat{\tau}_3\hat{\mathcal{G}}_\sigma\bigg]\!=\!0.
\end{equation}
Here, $\sigma=\uparrow\downarrow~(\zeta_\sigma=\pm 1)$ denotes the spin band and $\hat{\mathcal{G}}_\sigma$ is a matrix in Gor'kov-Nambu space \textit{only}:
\begin{equation}\label{eqn:GF_Nambu}
    \hat{\mathcal{G}}_\sigma(\vec{R}, {\vec{p}_\sigma})=\left(
    \begin{array}{cc}
        \mathcal{G}_\sigma(\vec{R}, {\vec{p}_\sigma}) &  \mathcal{F}_\sigma(\vec{R}, {\vec{p}_\sigma}) \\
         \tilde{\mathcal{F}}_\sigma(\vec{R}, {\vec{p}_\sigma}) &  \tilde{\mathcal{G}}_\sigma(\vec{R}, {\vec{p}_\sigma})
    \end{array}{}
    \right).
\end{equation}
Note that the two spin bands in general have different densities of states, which will prove crucial for the anomalous and Josephson diode effects presented in Sec.~\ref{sec:Results}. {Additionally, various interacting terms can be included by appropriate self-energy functions, as we discuss in more detail below}.

Now, we apply the quasiclassical approximation. Setting the momentum equal to the Fermi momentum of the respective spin band, $\vec{p}_\sigma\to\vec{p}_{F\sigma}$, multiplying the entire expression by $i/\pi$, and integrating over the energy close to the corresponding Fermi surface bring us to the  \textit{spin-scalar Eilenberger equation}~\cite{Eilenberger1968}:
\begin{equation}\label{eqn:Eilenberger}
i\hbar \vec{v}_\sigma\!\cdot\!\bm{\nabla}\hat{g}_\sigma(\vec{R})+\big[i\omega_n\hat{\tau}_3+\zeta_\sigma\hbar\vec{v}_\sigma\!\cdot\!\vec{Z}(\vec{R})\hat{\tau}_3-\hat{\Sigma}_\sigma,\hat{g}_\sigma(\vec{R})\big].
\end{equation}
Here,
\begin{equation}\label{eqn:Quasiclassical_GF}
\hat{g}_\sigma(i\omega_n, {\vec{p}_\sigma},\vec{R})=\frac{i}{\pi}\oint d\xi_\sigma\hat{\tau}_3\hat{\mathcal{G}}_\sigma(\xi_\sigma;i\omega_n, {\vec{p}_\sigma},\vec{R})
\end{equation}
is the quasiclassical Gor'kov Green's function for local spin projection $\sigma$, $\vec{v}_\sigma$ is the Fermi velocity associated with the respective spin band $\sigma=\uparrow\downarrow$ (for compactness we omit "F" in the subscript), and $\vec{Z}(\vec{R})=(i/4)\Tr[\hat{\sigma}_3\hat{U}(\vec{R})\bm{\nabla}_\vec{R}\hat{U}^\dagger(\vec{R})]$ is an effective U(1) field which coincides up to a prefactor with the adiabatic spin gauge field introduced in Sec.~\ref{sec:Intro}. An important observation is that this term enters the Eilenberger equation similarly to the standard U(1) orbital field, with the key distinction that it is now coupled to the spin, not the charge. {The aforementioned interacting terms (e.g., electron-impurity scattering) can be introduced via self-energy functions, $\hat{\Sigma}_\sigma$. Note that depending on the nature of the interaction, the self-energy can involve only one spin band, $\hat\Sigma_\sigma=\hat\Sigma_\sigma(\hat{g}_\sigma)$, or mixing between the two spin bands, $\hat\Sigma_\sigma=\hat\Sigma_\sigma(\hat{g}_\sigma,\hat{g}_{\bar{\sigma}})$, where $\bar\sigma$ is the opposite of $\sigma=\uparrow\!/\!\downarrow$. The results in this paper are obtained for a ballistic system with no other interactions besides the exchange field in the conical magnet, therefore, $\hat{\Sigma}_\sigma=\hat0$.}

The quasiclassical Green's function is normalized, and here we take the convention $\hat{g}_\sigma^2=\hat{\tau}_0$. In addition, the tilde symbol ($\tilde{~}$) refers to the particle-hole conjugation operation, which in imaginary-time (Matsubara) formalism takes the form 
\begin{equation}\label{eqn:particle_hole}
    \tilde{q}(i\omega_n, {\vec{p}_\sigma},\vec{R})=q^\ast(i\omega_n,-{\vec{p}_\sigma},\vec{R}).
\end{equation}
Once known, the quasiclassical Gor'kov Green's function allows to express observables in a compact manner. For instance, the Josephson current is calculated as
\begin{equation}\label{eqn:current}
\bm{j}(\vec{R})\!=\!\frac{-ie\pi k_BT}{2}\!\sum_{\sigma,n}N_\sigma\!\Tr\expval{\vec{v}_\sigma\hat{\tau}_3\hat{g}_\sigma(i\omega_n, {\vec{p}_\sigma},\vec{R})}_{\sigma},
\end{equation}
where $N_{\sigma}$ is the spin-resolved normal density of states at the corresponding Fermi level and $\expval{\bullet}_{\sigma}$ stands for the averaging over the respective Fermi surface, $\mathrm{FS}_\sigma$,
\begin{eqnarray}
    &&\expval{\bullet}_\sigma = \frac{1}{N_\sigma}\int_\mathrm{FS_\sigma}\frac{d^2p_\sigma}{(2\pi\hbar)^3|\vec{v}_\sigma(\vec{p}_\sigma)|}(\bullet),\\
    &&N_\sigma = \int_\mathrm{FS_\sigma}\frac{d^2p_\sigma}{(2\pi\hbar)^3|\vec{v}_\sigma(\vec{p}_\sigma)|}.
\end{eqnarray}
For definiteness, the results presented in Sec.~\ref{sec:Results} are obtained for spherical $\mathrm{FS}_\sigma$ which corresponds to the standard parabolic dispersion relation.

\subsection{Riccati amplitudes and boundary conditions}
The normalization condition $\hat{g}^2=\hat{\tau}_0$ (for compactness, here we omit the spin index $\sigma$) allows to choose an appropriate parametrization for the Green's function. In this paper, we write
\begin{equation}\label{eqn:Riccati_GF}
    \hat{g} = 2\left(
    \begin{array}{cc}
      g   & f \\
     -\tilde{f}   & -\tilde{g}
    \end{array}
    \right) - \hat{\tau}_3,
\end{equation}
where $g=(1-\gamma\tilde{\gamma})^{-1}$ and $f=(1-\gamma\tilde{\gamma})^{-1}\gamma$ with $\gamma$ and $\tilde{\gamma}$ being  so-called coherence functions \cite{eschrigElectromagneticResponseVortex1999,eschrigDistributionFunctionsNonequilibrium2000,eschrigScatteringProblemNonequilibrium2009},
which fulfill Riccati differential equations \cite{schopohlTransformationEilenbergerEquations1998}.

In accordance with the general theory, the $\gamma$ and $\tilde{\gamma}$ are related via the particle-hole conjugation operation~\eqref{eqn:particle_hole}. Therefore, it is sufficient to consider only, e.g., the $\gamma$ function, which satisfies the following linear differential equation:
\begin{equation}\label{eqn:Ricatti_F}
  i\hbar\vec{v}_\sigma\cdot(\bm{\nabla}-2i\zeta_\sigma\vec{Z})\gamma_\sigma + 2i\omega_n\gamma_\sigma=0,  
\end{equation}
with the formal solution 
\begin{equation}\label{eqn:gamma_F}
    \gamma_\sigma(i\omega_n,z)=A_\sigma e^{2i\zeta_\sigma\int\limits_a^z Z(\lambda)d\lambda} e^{-2\omega_nz/\hbar v_\sigma}.
\end{equation}
Note that the coordinate system is chosen such that $\vec{v}_\sigma=v_\sigma\vec{e}_z$ and $A_\sigma$ is the integration constant to be determined from the boundary conditions. In the superconducting state, the $\gamma_s$ coherence function is a matrix in spin space obeying the following Riccati differential equation,
\begin{equation}\label{eqn:Riccati_S}
    (i\hbar\vec{v}_F\cdot\bm{\nabla}+2i\omega_n)\hat{\gamma}_s + \hat{\Delta} - \hat{\gamma}_s\tilde{\Delta}\hat{\gamma}_s=0,
\end{equation}
whose solution for a homogeneous BCS (spin-singlet) state, $\bm{\nabla}\hat{\gamma}_s=\hat{0}$, reads
\begin{equation}\label{eqn:gamma_S}
    \hat{\gamma}_s(i\omega_n) = \frac{-\Delta\hat{\sigma}_2}{\omega_n+\sqrt{\omega_n^2+|\Delta|^2}}.
\end{equation}  
As indicated above, the $\tilde{\gamma}$ functions can be deduced from the corresponding $\gamma$ amplitudes arriving at 
\begin{eqnarray}
    \label{eqn:gammaTilde_F}
    &&\tilde{\gamma}_\sigma(i\omega_n,z) = \Tilde{A}_\sigma e^{-2i\zeta_\sigma\int\limits_a^z Z(\lambda)d\lambda} e^{2\omega_nz/\hbar v_\sigma},\\
    \label{eqn:gammaTilde_S}
    &&\tilde{\hat{\gamma}}_s(i\omega_n)= \frac{\Delta^\ast\hat{\sigma}_2}{\omega_n+\sqrt{\omega_n^2+|\Delta|^2}}.
\end{eqnarray}
This parametrization is suitable for applying boundary conditions formulated in terms of the normal-state S-matrix, which connects the incoming and outgoing Bloch waves on the two sides of the SC/FM interface. {To implement such boundary conditions for our model, we follow Refs~\cite{eschrigScatteringProblemNonequilibrium2009,greinSpinDependentCooperPair2009}, and for details we refer to Appendix~\ref{Appendix:BC}}.

\section{System under study}\label{sec:System}
\begin{figure}[t!]
    \centering
    \includegraphics[width=1\linewidth]{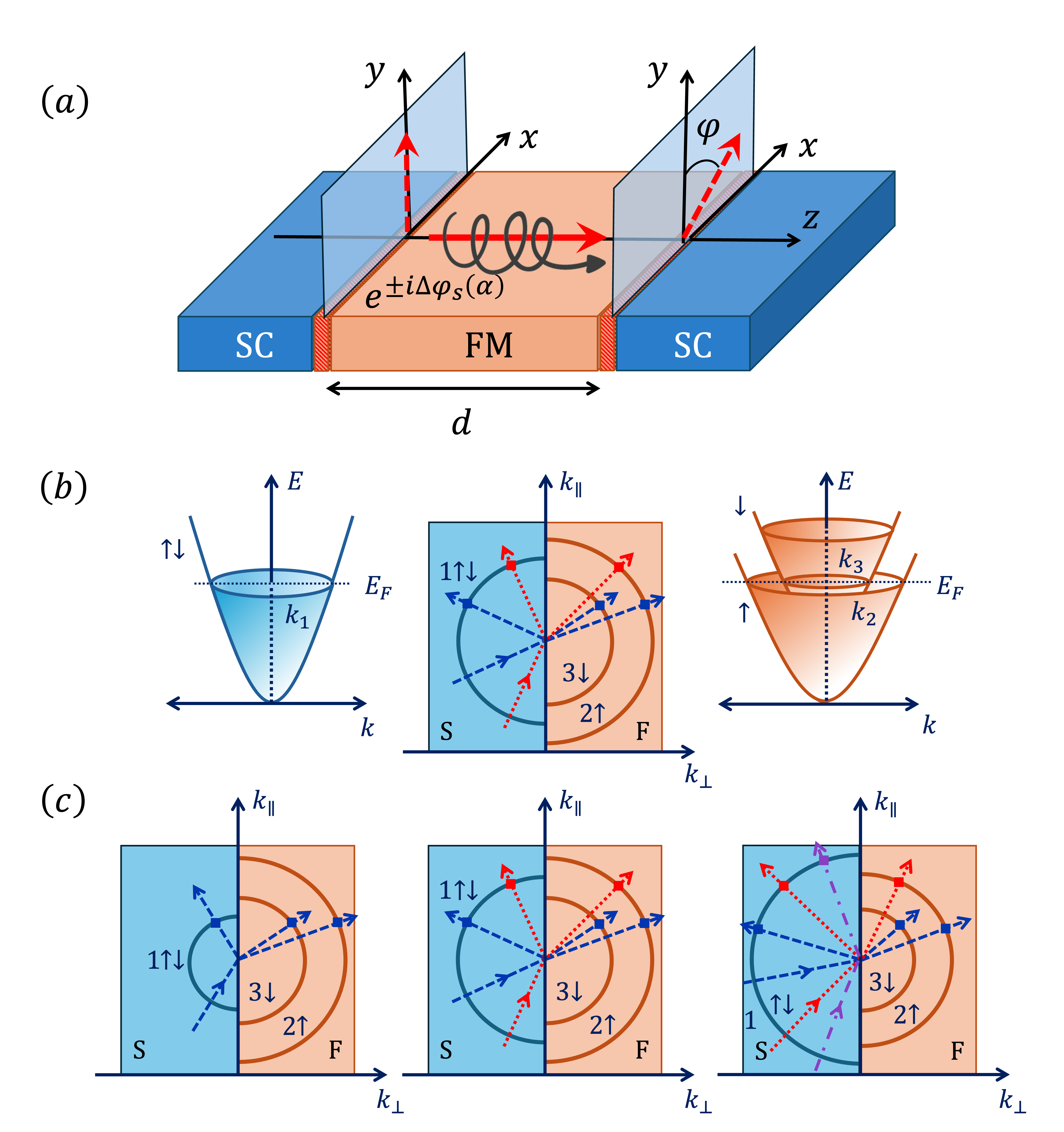}
   \caption{Panel (a): A scheme of the system under study where a strongly spin-polarized conical magnet (FM; orange) of length $d$ characterized by the magnetization profile $\vec{J}(z)$ in the adiabatic approximation (see the text below) is sandwiched between two identical BCS superconductors (SC; blue). The SF interfaces (shaded orange) are modeled by $\delta$-potentials allowing for nonvanishing magnetic moments pointing in an arbitrary direction. For clarity, and without loss of generality, we assume that the magnetization of the left interface is along the $y$ direction in the lab coordinate system ($z$-direction is the system's axis). The magnetic moment of the right interface lies in the $x-y$ plane and is characterized only by the azimuthal angle $\Delta\varphi$. Panel (b): Scheme of the SC/FM interface considered in this work. Thick lines show the corresponding Fermi surfaces on each side, which are assumed to be spherical. Panel (c): Possible Fermi surfaces geometries and allowed scattering processes (dotted, dashed, and dash-dotted lines) imposed by the conservation of the parallel component of momentum $\vec{k}_\parallel$.}
    \label{fig:System}
\end{figure}

We apply the model described above to a superconductor (SC) - conical magnet (FM) - superconductor system (SC) sketched in Fig.~\ref{fig:System}. The FM layer is depicted in the spirit of the adiabatic approximation, where the conical state can be locally viewed as an effective ferromagnetic state with the spin-$\uparrow$ and the spin-$\downarrow$ bands acquiring {different spin geometric phases described above}. The conical spin texture is given by
\begin{equation}
    \vec{J}(z) = J[\sin{\alpha}\cos{\phi(z)},\sin{\alpha}\sin{\phi(z)},\cos{\alpha}],
\end{equation}
where the spatial profile of the helix is determined by the $\phi(z)$ function. For simplicity, we assume a linear modulation, i.e., $\phi(z)=qz$, where $q$ relates to the pitch of the helix as $2\pi/q$. This leads to $\vec{Z}=-(1/4)(1-\cos{\alpha})\partial_z\phi\vec{e}_z=-(1/4)(1-\cos{\alpha})q\vec{e}_z$~[see Eq.~\eqref{eqn:Eilenberger}] and consequently the solutions for the coherence functions [see Eqs.~\eqref{eqn:gamma_F} and \eqref{eqn:gammaTilde_F}] read
\begin{eqnarray}
    \gamma_\sigma(z)= A_\sigma e^{i\zeta_\sigma\delta\varphi_s(\alpha, z)}e^{-2\omega_nz/\hbar v_\sigma},\\
    \tilde{\gamma}_\sigma(z)=\tilde{A}_\sigma e^{-i\zeta_\sigma\delta\varphi_s(\alpha, z)}e^{2\omega_nz/\hbar v_\sigma},
\end{eqnarray}
where $\delta\varphi_s(\alpha,z)=-(1/2)(1-\cos\alpha)qz$. Moreover, for simplicity and without loss of generality of the approach, we assume that the pitch of the helix equals the length of the helimagnet, i.e., $q=2\pi/d\implies\delta\varphi_s(\alpha,z)=-(1-\cos\alpha)(z/d)\pi$. Note that the above description can be straightforwardly generalized to the case of a spatially dependent $\alpha$. As anticipated in the previous section, the SC leads are assumed to be bulk homogeneous BCS superconductors characterized by Eqs.~\eqref{eqn:gamma_S} and \eqref{eqn:gammaTilde_S}. The symmetry discussion of the conical state is given in Appendix~\ref{Appendix:Symmetry}.

The remaining ingredient for obtaining a full description of the system is the information about the spin-polarized SC/FM interfaces, which are modeled by spin-dependent $\delta$-potentials, $U(z)=(V_{i}-J_i\vec{n}_i\cdot\bm{\hat{\sigma}})\delta(z-z_i)$. Here, $i=L,R$ refers to the left/right interface $(z_i=0,d)$, $V_{i}$ is the strength of the potential, $J_i$ is the spin splitting, and $\vec{n}_i=(\sin\alpha_i\cos\varphi_i,\sin\alpha_i\sin\varphi_i,\cos\alpha_i)$ is a unit vector referring to the direction of the magnetic moment in the respective barrier. For our model, it is crucial to have spin-polarized barriers that generate long-range equal-spin triplet correlations. However, unlike the case of a homogeneous FM~\cite{greinSpinDependentCooperPair2009,schulz2025_prl,schulz2025_prb}, the misalignment between $\vec{n}_L$ and $\vec{n}_R$ is not demanded due to the intrinsic noncoplanarity of a conical spin texture. In our calculations, we set $\alpha_L=\alpha_R=\pi/2$, and $\varphi_L=0$ and $\varphi_R=\Delta\varphi$ [see Fig.~\ref{fig:System}(a)]. Since the boundary conditions are evaluated at $z=0$ and $z=d$, this gives rise to the following spin geometric phase $\Delta\varphi_s =-(1-\cos\alpha)\pi$. 

As already mentioned, the SC/FM interfaces are characterized by the S-matrix, which, in general, has the following $4\times 4$ matrix form:
\begin{equation}\label{eqn:S_matrix}
    \bm{S} = \left(\begin{array}{ccc}
       \hat{R}_1 & \vv{T}_{12} & \vv{T}_{13} \\
        \vv{T}^T_{21} &  r_{22} & r_{23} \\
        \vv{T}^T_{31} &  r_{32} & r_{33}
    \end{array}\right).
\end{equation}
This structure of the S-matrix follows from the fact that the FM is characterized by two spin channels, whereas the SC by one doubly spin-degenerated channel. The $\hat{R}$ block describes the reflection (normal and spin-flip) in the superconducting side, while $r_{ij}$ amplitudes correspond to reflections in the ferromagnetic lead. Finally, $\vv{T}_{ij}=(t_{ij},t_{ij}^\prime)$ denotes the transmission amplitudes. The calculation of the S-matrix in the case of a spin-dependent $\delta$ potential introduced previously is straightforward, and we do not show the details here. We follow the notation in which the scattering channel of SC is labeled by $\eta=1$ and the two channels of FM by $\eta=2,3$ (corresponding to $\sigma=\uparrow\downarrow$ respectively) [see Figs.~\ref{fig:System}(b) and~\ref{fig:System}(c)]. In the context of spin-resolved currents in the FM layer, $\eta$ and $\sigma$ are used interchangeably.

\section{Results}\label{sec:Results}
In the following, we present the analytic and numerical results for the spin-resolved Josephson current phase-relations (CPR). All superconducting energy scales are expressed in the units of the superconducting gap at zero temperature, $\Delta_0$, length scales in the units of the superconducting coherence gap, $\xi=\hbar v_F/(2\Delta_0)$, and temperature in the units of the superconducting transition temperature, $T_c$. In contrast,
the energies of spin bands in the FM layer, $E_\eta$, are expressed in units of the Fermi energy in the SC leads, $E_F$. Finally, given the fact that we model the interfaces as spin-dependent $\delta$-potentials, we characterize them by the dimensionless BTK-like parameters~\cite{BTK1982}, $\mathcal{V}_{i}=2mV_{i}/(\hbar^2k_F)$ and $\mathcal{J}_{i}=2mJ_i/(\hbar^2k_F)$, where $k_F$ denotes the Fermi wave vector in the SC. The superconductors are assumed to be identical up to the superconducting phase, and most of the results are obtained for $T/T_c=0.1$, $d/\xi=0.4$, and $\alpha_L=\alpha_R=\pi/2$ if not stated otherwise. 
\subsection{Tunneling limit: "$\varphi_0$-junction"}
The tunneling limit is realized in the case of strong interfacial barriers between the SC and the FM regions, characterized by $\mathcal{V}_{i}\gg 1$. For simplicity, the analytic consideration below will be given for the perpendicular (single-trajectory) impact on the SC/FM interfaces. The full numerical result that confirms our conclusions will be given afterward. 

The main assumption is that the tunnel barriers imply small coherence amplitudes allowing us to linearize the boundary conditions in $\gamma_\eta$ and $\tilde{\gamma}_\eta$ and the transmission coefficient and treat the problem analytically. As shown in Appendix~\ref{Appendix:Linearized_BC}, the outgoing amplitudes at the two SC/FM interfaces, $\Gamma_{\eta j}$ and $\tilde{\Gamma}_{\eta j}~(j=L,R)$, can be calculated from the following linear system of equations~\cite{greinSpinDependentCooperPair2009}: 
\begin{equation}\label{eqn:Gamma}
\begin{split}
    \vv{\Gamma}_L &\,= \hat{M}_L\hat{\beta}^\ast\vv{\Gamma}_R + \vv{A}_L,\\
    \vv{\Gamma}_R &\,= \hat{M}_R\hat{\beta}\vv{\Gamma}_L + \vv{A}_R,
\end{split}
\end{equation}
where
\begin{equation}
    \begin{split}
    &\vv{\Gamma}_j =\left(\begin{array}{cc}
       \Gamma_2 \\
       \Gamma_3 
    \end{array}\right)^j~,~\vv{A}_j =\left(\begin{array}{cc}
       A_2 \\
       A_3 
    \end{array}\right)^j,\\
    \hat{M}_j=&\left(\begin{array}{cc}
      |r_{22}|^2  & r_{23}r^\ast_{32} \\
      r_{32}r^\ast_{23}   & |r_{33}|^2
   \end{array}\right)^j~,~
   \hat{\beta} = \left(\begin{array}{cc}
      \beta_2  & 0\\
      0   & \beta_3
   \end{array}\right).
   \end{split}
\end{equation}
Here, $r_{\eta\eta^\prime}$ are introduced above [see Eq.~\eqref{eqn:S_matrix}], $\beta_\eta=e^{i\zeta_\eta\Delta\varphi_s}e^{-2|\omega_n| d/\hbar v_\eta}$ with $\zeta_\eta=\pm 1$ for $\eta=2,3$, and
\begin{equation}
    \label{eqn:A}
    A_\eta\!=\! \vv{T}_{\eta1}^{T}\qty[\hat{\gamma}_1\!+\!\hat{\gamma}_1\hat{R}^\ast_1(1\!-\!\tilde{\hat{\gamma}}_1\hat{R}_1\hat{\gamma}_1\hat{R}^\ast_1)^{-1}\tilde{\hat{\gamma}}_1\hat{R}_1\hat{\gamma}_1]\vv{T}_{1\eta}^\ast.
\end{equation}
Taking the above details into account, Eqs.~\eqref{eqn:Gamma} can be easily inverted, arriving at the following solutions:
\begin{align}
\vv{\Gamma}_L &\,= \left(1-\hat{M}_L\hat{\beta}^\ast\hat{M}_R\hat{\beta}\right)^{-1}\left(\hat{M}_L\hat{\beta}^\ast\vv{A}_R+\vv{A}_L\right),\\
\vv{\Gamma}_R &\,= \left(1-\hat{M}_R\hat{\beta}\hat{M}_L\hat{\beta}^\ast\right)^{-1}\left(\hat{M}_R\hat{\beta}\vv{A}_L+\vv{A}_R\right).
\end{align}

To obtain the Josephson current in this regime, one should linearize Eq.~\eqref{eqn:current} as well:
\begin{eqnarray}
    \bm{j}(\vec{R}) &&\,= \frac{-ie\pi k_BT}{2}\sum_{\eta}N_\eta\sum_{n}\expval{\vec{v}_\eta\frac{1+\gamma_\eta(i\omega_n)\tilde{\gamma}_\eta(i\omega_n)}{1-\gamma_\eta(i\omega_n)\tilde{\gamma}_\eta(i\omega_n)}}_{\eta}\nonumber\\
    &&\,=-ie\pi k_BT\sum_{\eta}N_\eta\sum_{n}\sum_m\expval{\vec{v}_\eta\left[\gamma_\eta(i\omega_n)\tilde{\gamma}_\eta(i\omega_n)\right]^m}_\eta\nonumber\\
    \label{eqn:Josephson_tunneling}
    &&\,\approx -iek_B\pi T\sum_{\eta}N_\eta\sum_{n}\expval{\vec{v}_\eta\gamma_\eta(i\omega_n)\tilde{\gamma}_\eta(i\omega_n)}_\eta.
\end{eqnarray}
In the last step, we have employed the fact that $\gamma$ and $\tilde{\gamma}$ are small in the FM due to the weak proximity effect (strong barriers at the interfaces). The above formula can be used to calculate the Josephson CPR in the SC/FM/SC system under consideration
\begin{eqnarray}\label{eqn:current_tunneling}
    \bm{j}(z=0)&&\,=-ie\pi k_BT\times\\
    &&\,\times \sum_\eta N_\eta\sum_n \expval{\vec{v}_\eta [\Gamma_{\eta L}\beta_\eta\Gamma_{\eta R}^\ast-\Gamma_{\eta L}^\ast\beta_\eta^\ast\Gamma_{\eta R}]}_{\eta+}.\nonumber
\end{eqnarray}
Note that due to the current conservation, we evaluate it at the left SC/FM interface, which is at $z=0$, and that the averaging is performed only over the positive directions of the momentum, i.e., $p_z>0$, for each spin band. The same expression would appear if we considered the right SC/FM interface at $z=d$, instead.  
\begin{figure}[t!]
    \centering
    \includegraphics[width=1\linewidth]{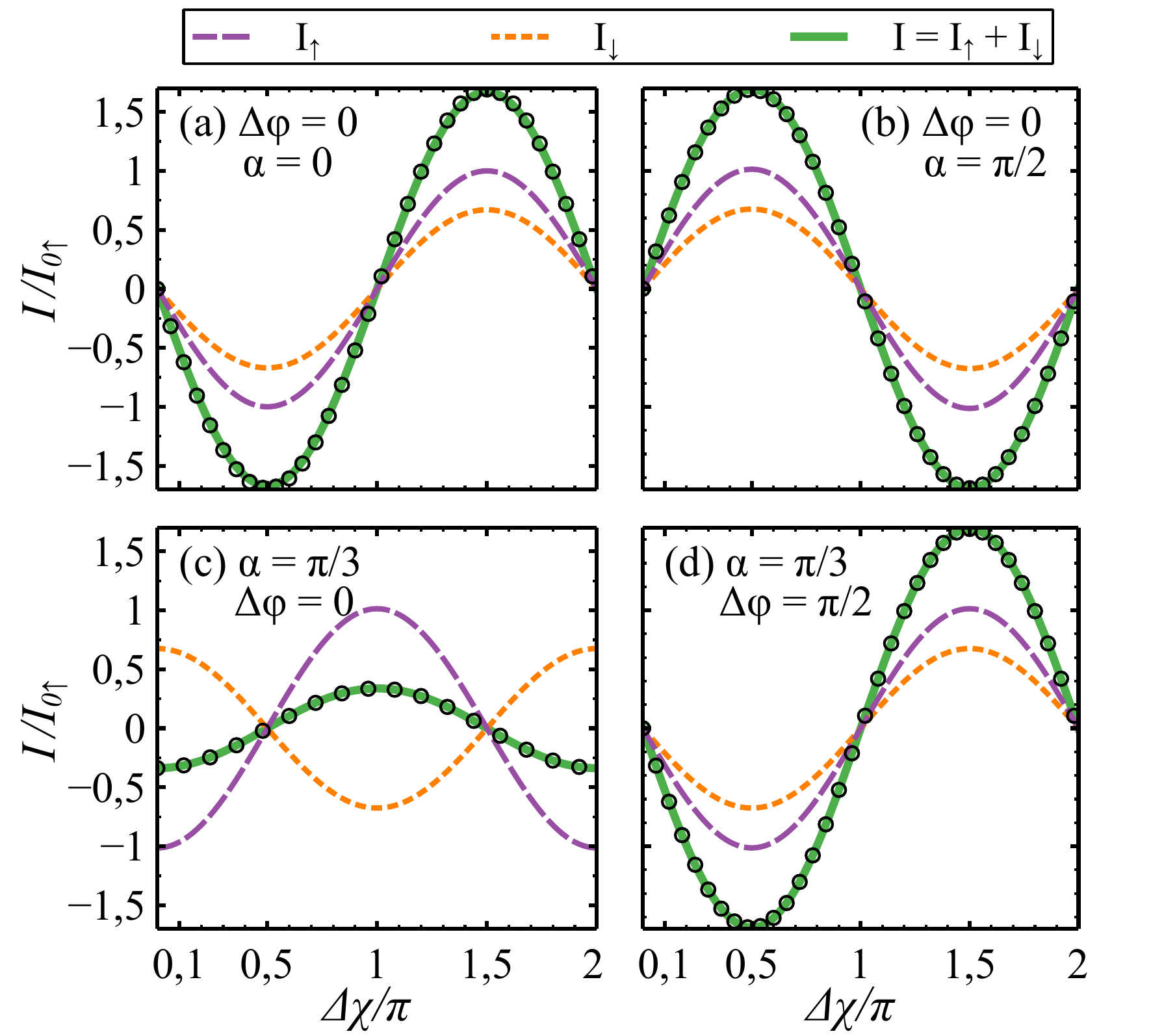}
    \caption{The Josephson CPR in an SC/FM/SC junction for different $\alpha$ and $\Delta\varphi$ angles in the tunneling limit calculated from Eq.~\eqref{eqn:current_tunneling}. The junction's parameters are $E_2=1.9E_F$, $E_3=1.2E_F$, $\mathcal{V}_L=\mathcal{V}_R=50$, $\mathcal{J}_L=\mathcal{J}_R=1$, $d=0.4\xi$, and $T=0.1T_c$. The dashed and dotted lines represent the spin-resolved supercurrents, while the solid green lines correspond to the total charge Josephson current. The black circles denote the full numerical solution based on the iteration procedure [see Sec.~\ref{sec:JDE}].}
    \label{fig:CPR_T}
\end{figure}

From Eq.~\eqref{eqn:A} it can be shown that $A_{\eta j}\propto e^{i(\chi_j-\zeta_\eta\varphi_j)}$, $(j=L,R)$. Evaluating the current~\eqref{eqn:current_tunneling} to the leading order and restricting ourselves to the single trajectory case, $\vec{p}=p\vec{e}_z$, we obtain
\begin{eqnarray}
        j_\eta &&\,\propto\\
        &&\,\propto\!\Im[\beta_\eta A_{\eta L} A^\ast_{\eta R}] \propto \Im[e^{i\zeta_\eta\Delta\varphi_s}e^{i(\chi_L-\zeta_\eta\varphi_L)}e^{-i(\chi_R-\zeta_\eta\varphi_R)}]\nonumber\\
        &&\,\!=\!\Im e^{-i(\Delta\chi-\zeta_\eta\Delta\varphi_s-\zeta_\eta\Delta\varphi)}\!=\!-\sin(\Delta\chi-\zeta_\eta\Delta\varphi_s-\zeta_\eta\Delta\varphi)\nonumber,
\end{eqnarray}
where $\Delta\chi=\chi_R-\chi_L$ and $\Delta\varphi=\varphi_R-\varphi_L$. In other words, 
\begin{equation}\label{eqn:CPR_tunneling}
    I_\eta = - I_{0\eta}\sin(\Delta\chi-\zeta_\eta\Delta\varphi_s-\zeta_\eta\Delta\varphi),
\end{equation}
 and the equilibrium phase shift is present as long as $I_{02}\neq I_{03}$. Although sinusoidal the Josephson CPR is shifted by two contributions: (i) geometric phase shift $\Delta\varphi$ due to the noncoplanar configuration of the magnetizations in the ferromagnetic trilayer and (ii) the spin geometric phase shift $\Delta\varphi_s-(1-\cos\alpha)\pi$ due to the intrinsic noncoplanarity of the conical state in the FM region. Consequently, here we draw three important conclusions. First, a homogeneous state obtained for $\alpha=0$ displays no anomalous Josephson effect, $\Delta\varphi_s(\alpha=0)=0$, as long as $\Delta\varphi=n\pi$, $n\in \mathbb{Z}$ {and the junction is in the $\pi$ state}. Second, in-plane helical ordering, $\alpha=\pi/2$, leads to a $\pi-0$ transition, $\Delta\varphi_s(\alpha=\pi/2)=-\pi\implies I_\eta\to-I_\eta$. Third, the effect vanishes in the case of $\Delta\varphi+\Delta\varphi_s= n\pi, n\in\mathbb{Z}$. 
  
  As indicated, the analytic discussion above was done for a single-trajectory case. To consider all trajectories, one numerically evaluates Eq.~\eqref{eqn:current_tunneling}, and the resulting Josephson CPRs are shown in Fig.~\ref{fig:CPR_T}. The dashed and dotted curves represent the spin-resolved Josephson currents, while the solid green line corresponds to the total charge current. The CPRs are presented for various values $\alpha$ and $\Delta\varphi$, and the tunneling limit is realized by setting $\mathcal{V}_i=50$ with other parameters indicated in the caption. Panel (a) shows the CPRs for a homogeneous FM, $\alpha=0$, and the parallel interface monetizations, $\Delta\varphi=0$. The normal Josephson effect is observed and the junction is in the $\pi$ state. Examining the in-plane helical state in the FM, $\alpha=\pi/2$, leads to the 0-junction, as shown in panel (b). It is important to note that in both cases, the junction is inversion symmetric, and the normal Josephson effect is implied. In contrast, a conical state of the FM results in the anomalous Josephson effect or $\varphi_0$-junction, as seen in panel (c), which displays the CPRs for $\alpha=\pi/2$ and parallel interface magnetizations, $\Delta\varphi=0$. The phase shift of the charge Josephson current (green line) is entirely determined by $\alpha$ and amounts to $\varphi_0=-(1-\cos\alpha)\pi=-\pi/2$. Finally, the normal junction can be realized by $\Delta\varphi-(1-\cos\alpha)\pi=n\pi, n\in\mathbb{Z}$. This situation is presented in Fig.~\ref{fig:CPR_T}(d) where the CPRs are calculated for $\alpha=\pi/3$ and $\Delta\varphi=\pi/2$. In all panels, the black circles indicate the full numerical solution based on the iteration procedure outlined in the following subsection, justifying the approximations from Eqs.~\eqref{eqn:Gamma}-\eqref{eqn:Josephson_tunneling}. Note that all conclusions we have drawn from the single-trajectory case remain valid here.
  
\subsection{Josephson diode effect}\label{sec:JDE}
As anticipated in the previous section, to have a JDE we need to go beyond the tunneling limit. To account for higher harmonics in the CPR, the boundary condition problem [see Appendix~\ref{Appendix:BC}] should be solved iteratively at the two SC/FM interfaces. Namely, the outgoing amplitudes at one interface propagate throughout the FM and approach the other interface as the incoming amplitudes. The convergence is ensured by the fact that each next iteration is suppressed by a factor of $\beta_\eta=e^{i\zeta_\eta\Delta\varphi_s}e^{-2|\omega_n| d/v_\eta}$ as compared to the previous one.

As already mentioned, the figure of merit in this effect is the so-called diode efficiency, defined as
\begin{equation}
    \eta_\mathrm{e}(\alpha,\Delta\varphi)= \frac{|I^+|-|I^-|}{|I^+|+|I^-|},
\end{equation}
where $I^\pm \equiv I(\Delta\chi_\pm)$ with $\Delta\chi_\pm$ being the superconducting phase differences at which the positive $(+)$ or negative $(-)$ critical charge current is reached for fixed $\alpha$ and $\Delta\varphi$, i.e., $\Delta\chi^+ = \mathrm{argmax}_{\Delta\chi}(I)$ and $\Delta\chi^- = \mathrm{argmin}_{\Delta\chi}(I)$. The role of $\Delta\varphi$ has been thoroughly examined in our Refs.~\cite{schulz2025_prl,schulz2025_prb}. Therefore, here we mainly focus on the spin geometric phase $\Delta\varphi_s$. We show that intrinsically noncoplanar spin textures can yield a prominent JDE even in the case of parallel magnetic moments in the barriers, $\Delta\varphi=0$. Note that the spin-polarized interfaces are needed to generate equal-spin triplet correlations necessary for the effect in question. Therefore, in this subsection, we mainly focus on the role of $\alpha$ while the other parameters are $E_2=1.2E_F$, $E_3=0.4E_F$, $\mathcal{V}_{i}=0.2$, and $\mathcal{J}_i=0.1$ ($i=L, R$). The current is normalized to $\Delta_0\pi/(eR_N)$ where $R_N$ is the resistance of the junction in the normal state with collinear magnetizations of the three magnetic layers {$(\alpha_L=\alpha_R=0)$}
\begin{equation}
    \frac{1}{R_N\mathcal{A}} = e^2\sum_{\eta} N_{\eta}\expval{{v}_{\eta z}(\vec{p})\frac{\mathcal{T}^L_\eta(\vec{p})\mathcal{T}^R_\eta(\vec{p})}{1+\mathcal{R}^L_\eta(\vec{p})\mathcal{R}^R_\eta(\vec{p})}}_{\eta+}.
\end{equation}
Here, $\mathcal{T}^{L/R}_\eta=|t^{L/R}_{1\eta}|^2$ ($\mathcal{R}^{j}_\eta=1-\mathcal{T}^j_\eta$) is the transmission probability for an electron from band $\eta$ in the FM into the left/right normal lead [see Eq.~\eqref{eqn:S_matrix}] and $\mathcal{A}$ is the cross-section of the junction. Note that the above formula is valid only in the case which excludes spin-flip processes, and for this reason, we demand {$\alpha_L=\alpha_R=0$}.  

\begin{figure}[t!]
    \centering
    \includegraphics[width=1\linewidth]{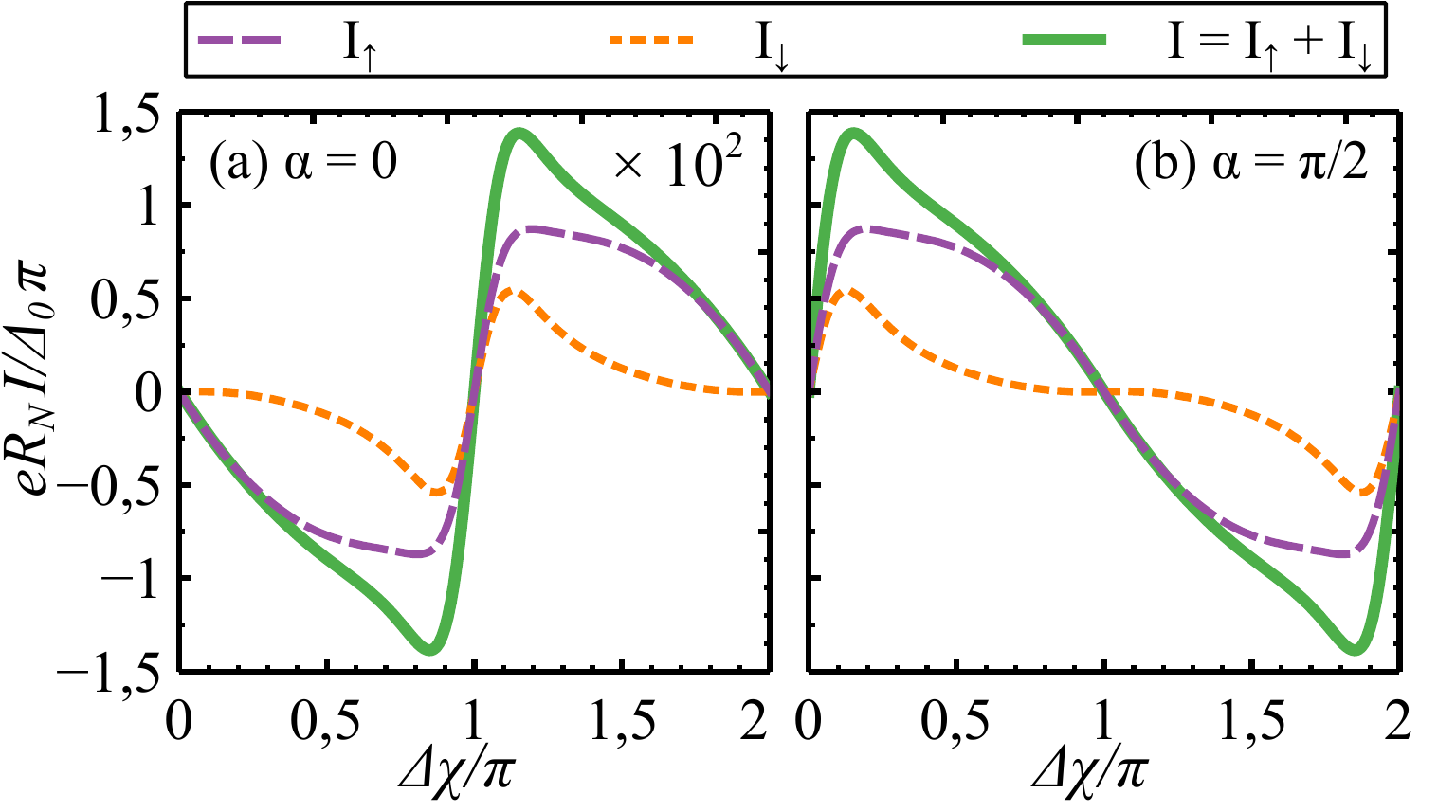}
    \caption{The spin-resolved Josephson CPR across the FM layer of length $d=0.4\xi$, energies of the spin bands $E_2=1.2 E_F$ and $E_3=0.4E_F$, and (a) $\alpha=0$ (homogeneous ferromagnet) and (b) $\alpha=\pi/2$ (in-plane helical state). Other parameters are $\mathcal{V}_L=\mathcal{V}_R=0.2$, $\mathcal{J}_L=\mathcal{J}_R=0.1$, $\alpha_L=\alpha_R=\pi/2$, $\Delta\varphi=0$, and $T=0.1T_c$.}
    \label{fig:CPR_normal}
\end{figure}

Let us first examine the case of junctions with coplanar spin textures that possess inversion symmetry. Figure~\ref{fig:CPR_normal} displays the Josephson CPR for the parameters mentioned above, considering a homogeneous ferromagnet [$\alpha=0$; panel (a)] and an in-plane helical magnet [$\alpha=\pi/2$; panel (b)]. As before, the dashed and dotted lines represent the spin-resolved Josephson currents, while the solid green line corresponds to the total charge current. Due to the inversion symmetry, both cases display the normal Josephson effect, $I_\eta(-\Delta\chi)=-I_\eta(\Delta\chi)$. However, we note that, similarly to the tunneling case, the homogeneous ferromagnetic state features the $\pi$ phase, while the helical state features the $0$ phase.

\begin{figure}[t!]
    \centering
    \includegraphics[width=1\linewidth]{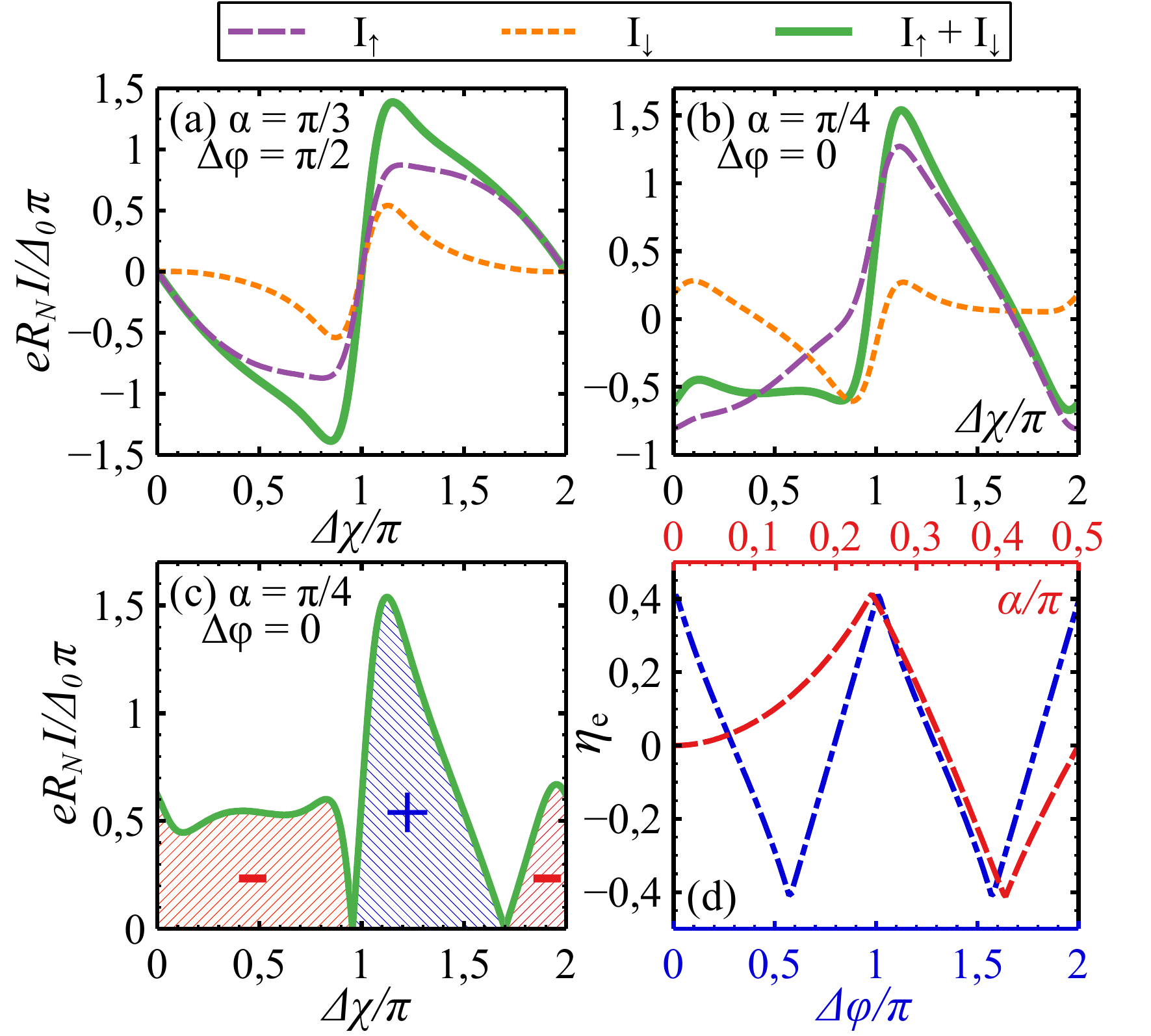}
    \caption{The spin-resolved Josephson CPR in a ferromagnetic trilayer for (a) $\alpha=\pi/3$ and $\Delta\varphi=\pi/2$ and (b) $\alpha=\pi/4$ and $\Delta\varphi=0$. Other parameters are as in Fig.~\ref{fig:CPR_normal}. Panel (c): Absolute value of the total Josephson current from panel (b) showing the Josephson diode effect explicitly. The red and blue areas display the negative and the positive currents, respectively. Panel (d): The diode efficiency $\eta_\mathrm{e}$ as a function of $\alpha$ for $\Delta\varphi=0$ (red dashed line) and $\Delta\varphi$ for $\alpha=\pi/3$ (blue dashed-dotted line).}
    \label{fig:JDE}
\end{figure}

Let us now consider a noncoplanar spin texture. While this state is expected to exhibit a JDE, we first demonstrate that this is not always true. As discussed in the previous subsection, the condition $\Delta\varphi+\Delta\varphi_s = 0 \implies \Delta\varphi = (1- \cos\alpha)\pi$ yields no anomalous Josephson effect in the tunneling limit [see Fig.~\ref{fig:CPR_T}(d)]. Here, we establish that a similar scenario applies in the case of arbitrary transparency. Figure~\ref{fig:JDE}(a) displays the spin-resolved Josephson CPR for $\alpha=\frac{\pi}{3}$ and $\Delta\varphi=\frac{\pi}{2}$, with all other parameters being the same as in Fig.~\ref{fig:CPR_normal}. We note that both the spin-resolved (dashed and dotted lines) and the total charge CPR (solid green line) demonstrate a normal Josephson effect since $\Delta\varphi=(1- \cos\alpha)\pi$. To understand why this condition results in a normal Josephson effect, we introduce an effective phase $\Delta\varphi^\prime=\Delta\varphi+\Delta\varphi_s = \Delta\varphi-(1- \cos\alpha)\pi$ which, henceforth, we term as \textit{the total spin geometric phase}. Keeping in mind that $\alpha$ and $\Delta\varphi$ under the inversion operation transform as $\alpha\to\pi-\alpha$ and $\Delta\varphi\to 2\pi-\Delta\varphi$, this phase transforms as $\Delta\varphi^\prime=\Delta\varphi-(1-\cos\alpha)\pi\to 2\pi-\Delta\varphi-[1-\cos(\pi-\alpha)]\pi=2\pi-\Delta\varphi-(1+\cos\alpha)\pi=-\Delta\varphi+(1-\cos\alpha)\pi=-\Delta\varphi^\prime$. Consequently, only the scenario of $\Delta\varphi^\prime=0 \Leftrightarrow \Delta\varphi=\Delta\varphi_s=(1-\cos\alpha)\pi$ possesses inversion symmetry, which leads to the normal Josephson effect. As will be demonstrated below, this result leads to the necessary conditions for the occurrence of the JDE: (i) a noncoplanar spin texture of the ferromagnetic trilayer, (ii) different density of states for the two spin bands in the FM layer, and (iii) $\Delta\varphi^\prime\neq k\frac{\pi}{2}$, where $k$ is an integer. The latter condition can be understood by the fact that $\pm\Delta\varphi^\prime$ is the total phase acquired by $\uparrow\uparrow$- and $\downarrow\downarrow$-correlations, respectively. Therefore, these phase factors either coincide for $\Delta\varphi^\prime=k\pi$ or differ by multiplies of $\pi$ for $\Delta\varphi^\prime=\frac{\pi}{2}+k\pi$. Note that in the case of a homogeneous state, $\alpha = 0$, this condition reduces to $\Delta\varphi\neq k\frac{\pi}{2}$, as discussed in detail in Ref.~\cite{schulz2025_prl,schulz2025_prb}. On the other hand, if $\Delta\varphi = 0 $, the condition simplifies to $(1-\cos\alpha)\neq \frac{k}{2}\implies \alpha\neq 0, \frac{\pi}{3}, \text{and}~\frac{\pi}{2}$.

Figure~\ref{fig:JDE}(b) shows the spin-resolved Josephson CPR for $\alpha=\frac{\pi}{4}$ and $\Delta\varphi=0$, other parameters are the same as in panel (a). Due to the noncoplanarity and $(1-\cos\alpha)\neq \frac{k}{2}$, the strong asymmetries between the positive and negative critical currents in all CPRs emerge. For better visibility, the absolute value of the total charge current (solid green line) is presented in panel (c), where the red and blue areas show the negative and the positive currents, respectively. The appearance of the JDE is evident. To quantify it, in Fig.~\ref{fig:JDE}(d) we present the diode efficiency as a function of $\alpha$ for $\Delta\varphi=0$ (dashed red line) and $\Delta\varphi$ for $\alpha=\frac{\pi}{3}$ (dash-dotted blue line). In both cases, the function displays a non-monotonic behavior exceeding the value of $\eta_\mathrm{e}\approx 41\%$. Let us discuss first the role of $\alpha$. Since $\Delta\varphi=0$ we immediately notice the absence of the JDE for $\alpha_0=0,\frac{\pi}{3}, \text{and}~\frac{\pi}{2}$, which is in accordance with the discussion above. The maximum values of $\eta_\mathrm{e}\approx \pm 41\%$ are achieved for $\alpha\approx 0.25\pi$ and $\alpha\approx 0.4\pi$, respectively. The role of $\Delta\varphi$ is similar. The JDE is absent for $\Delta\varphi^\prime_0=\Delta\varphi_0-\frac{\pi}{2} = k\frac{\pi}{2}\implies \Delta\varphi_0 = 0,\frac{\pi}{2},\frac{3\pi}{2}, \text{and}~2\pi$. On the other hand, the maximum values of $\eta_\mathrm{e}\approx \pm 41\%$ are achieved for $\Delta\varphi\approx 0.22\pi + k\pi$ and $\Delta\varphi\approx 0.78\pi + k\pi$, respectively.  

\subsection{Harmonic analysis}\label{sec:Harmonic}
As already indicated, higher harmonics in the Josephson current-phase relation are crucial for the Josephson diode effect. In addition, we stress the importance of two quantum phases that add to the superconducting phase difference $\Delta\chi$: (i) the adiabatic spin geometric phase $\Delta\varphi_s=-(1-\cos{\alpha})\pi$ and (ii) the quantum-geometric phase difference $\Delta\varphi$. As discussed above, and as we show in Appendix~\ref{Appendix:Fourier}, these two can be joined into the \textit{total spin geometric phase} defined as $\Delta\varphi^\prime=\Delta\varphi+\Delta\varphi_s=\Delta\varphi-(1-\cos\alpha)\pi$. Under the time reversal, this phase transforms as $\Delta\varphi^\prime\to-\Delta\varphi^\prime$ laying the basis for the harmonic analysis presented in the following. Namely, under time reversal the equal-spin currents transform as
\begin{equation}\label{eqn:I_time_reversal}
    I_\eta(\Delta\chi, {\Delta\varphi^\prime})=-I_\eta(-\Delta\chi, {-\Delta\varphi^\prime}).
\end{equation}
As we show below, this relation leads to the following harmonic decomposition of the CPRs:
\begingroup
\allowdisplaybreaks
\begin{align}
    I_{\uparrow} &= \frac{1}{2}\sum_{\mu,\nu=-\infty}^\infty  (-1)^{\mu+\nu} \mu I_{\mu,\nu} \sin\psi_{\mu,\nu}, \label{eqn:I_uu} \\ 
    I_{\downarrow} &= \frac{1}{2}\sum_{\mu,\nu=-\infty}^\infty  (-1)^{\mu+\nu} \nu I_{\mu,\nu} \sin\psi_{\mu,\nu}, \label{eqn:I_dd}
\end{align}
\endgroup
where $I_{-\mu,-\nu}=I_{\mu,\nu}$, and 
\begin{equation}
\psi_{\mu,\nu} = (\mu+\nu)\Delta\chi - (\mu-\nu)\Delta\varphi^\prime \label{eqn:effJos}
\end{equation}
is an \textit{effective Josephson phase}. A similar form has been suggested for the first time in Ref.~\cite{eschrigSpinpolarizedSupercurrentsSpintronics2015} for ballistic systems featuring quantum-geometric phases $\Delta\varphi$. However, Eq.~\eqref{eqn:effJos} is modified due to the spin geometric phase $\Delta\varphi_s$.  {A similar expansion for in diffusive systems has been proposed in Refs.~\cite{schulz2025_prl,schulz2025_prb} accompanied by a physically intuitive picture, which we here just briefly outline. Namely, Eqs.~\eqref{eqn:I_uu} and \eqref{eqn:I_dd} can be understood as the CPRs mediated by a coherent transfer of $\mu$ $\uparrow\uparrow$ and $\nu$ $\downarrow\downarrow$ equal-spin pairs. Noting that the ground state for $\Delta\varphi^\prime=0$ is a $\pi$-junction [see Figs.~\ref{fig:CPR_normal}(a) and \ref{fig:JDE}(a)] and the Josephson phase acquired during a coherent transport of a single
 $\uparrow\uparrow\!/\!\downarrow\downarrow$ pair is $\Delta\chi \mp \Delta\varphi^\prime + \pi$, we end up with Eq.~\eqref{eqn:effJos}, which is a Josephson phase corresponding to the process that involve $\mu$ $\uparrow\uparrow$-pairs and $\nu$ $\downarrow\downarrow$-pairs.}

As mentioned above, the Fourier analysis mainly relies on Eq.~\eqref{eqn:I_time_reversal}, which allows for performing the following harmonic expansion:
\begin{align}
        I_{\eta}(\Delta \chi,\Delta \varphi') &= \sum_{m=1}^\infty 
        A_{m,0}^\eta \sin(m\Delta\chi) +
        \sum_{n=1}^\infty 
        B_{0,n}^\eta\sin(n{\Delta\varphi^\prime}) \nonumber \\
        &+\sum_{m,n=1}^\infty \big[A_{m,n}^\eta \sin(m\Delta\chi)\cos\qty(n{\Delta\varphi^\prime})
        \label{eqn:general_Fourier_ansatz}\\
        &\qquad \quad +  B_{m,n}^\eta \cos(m\Delta\chi)\sin\qty(n{\Delta\varphi^\prime})\big]. \nonumber
\end{align}
Here,
\begin{align}
    \label{eqn:Amn}
    &A^\eta_{m,n}=\int\limits_0^{2\pi}\frac{d\chi}{\pi}\int\limits_0^{2\pi}\frac{d\varphi^\prime}{\pi} I_\eta(\chi,\varphi^\prime)\sin(m\chi)\cos(n\varphi^\prime),\\
    \label{eqn:Bmn}
    &B^\eta_{m,n}=\int\limits_0^{2\pi}\frac{d\chi}{\pi}\int\limits_0^{2\pi}\frac{d\varphi^\prime}{\pi} I_\eta(\chi,\varphi^\prime)\cos(m\chi)\sin(n\varphi^\prime)
\end{align}
for $m,n\neq 0$, and
\begin{align}
    \label{eqn:Am0}
    &A^\eta_{m,0}=\int\limits_0^{2\pi} \frac{d\chi}{\pi}\int\limits_0^{2\pi} \frac{d\varphi^\prime}{2\pi}I_\eta(\chi,\varphi^\prime)\sin(m\chi),\\
    \label{eqn:B0n}
    &B^\eta_{0,n}\!=\!\int\limits_0^{2\pi} \frac{d\chi}{2\pi}\int\limits_0^{2\pi} \frac{d\varphi^\prime}{\pi}I_\eta(\chi,\varphi^\prime)\sin(n\varphi^\prime).
\end{align}
{Derivation of Eqs.~\eqref{eqn:general_Fourier_ansatz}-\eqref{eqn:B0n}  is given in Appendix~\ref{Appendix:Fourier}. {Based on the numerical treatment of the problem, we find that the spin-resolved currents depend only on $\Delta\chi$ and $\Delta\varphi^\prime$ leading to Eq.~\eqref{eqn:general_Fourier_ansatz}. This fact was already indicated in the analytic treatment of the tunneling limit [see Eq.~\eqref{eqn:CPR_tunneling}]. Once derived, Eq.~\eqref{eqn:general_Fourier_ansatz} implies Eqs.~\eqref{eqn:I_time_reversal}-\eqref{eqn:effJos}, as was shown in Refs.~\cite{schulz2025_prl,schulz2025_prb}.} 
In addition, we find that the $A_{m,n}$ and $B_{m,n}$ coefficients do not depend on $\Delta\varphi_s$, i.e., on the conical angle $\alpha$. On the other hand, they do depend on the polar angles characterizing the direction of barriers' magnetic moments, $\alpha_L$ and $\alpha_R$. However, since the case of $\alpha_L=\alpha_R=\pi/2$ maximizes the effect, we limit our discussion in this paper to this case, and do not scrutinize the effect of different values of $\alpha_L$ and $\alpha_R$ here. Finally, $A_{m,n}$ and $B_{m,n}$ are not completely independent, as we discuss below. For a more detailed derivation of the relations between them, see Ref.~\cite{schulz2025_prb}.} Particularly, we find that $A_{m,n}$ and $B_{m,n}$ are nonzero only when $m\pm n$ is even leading to the following relation between the coefficients in Eqs.~\eqref{eqn:I_uu} and~\eqref{eqn:I_dd} and Eq.~\eqref{eqn:general_Fourier_ansatz}:
\begin{align}
    \label{eqn:I_mu}
   (-1)^{\mu+\nu} |\mu| I_{\mu,\nu}=
    \frac{A^\uparrow_{|\mu+\nu|,|\mu-\nu|}-B^\uparrow_{|\mu+\nu|,|\mu-\nu|}}{2},\\
    \label{eqn:I_nu}
    (-1)^{\mu+\nu} |\nu| I_{\mu,\nu}=
    \frac{A^\downarrow_{|\mu+\nu|,|\mu-\nu|}-B^\downarrow_{|\mu+\nu|,|\mu-\nu|}}{2}.
\end{align}
\begin{figure}
    \centering
    \includegraphics[width=1\linewidth]{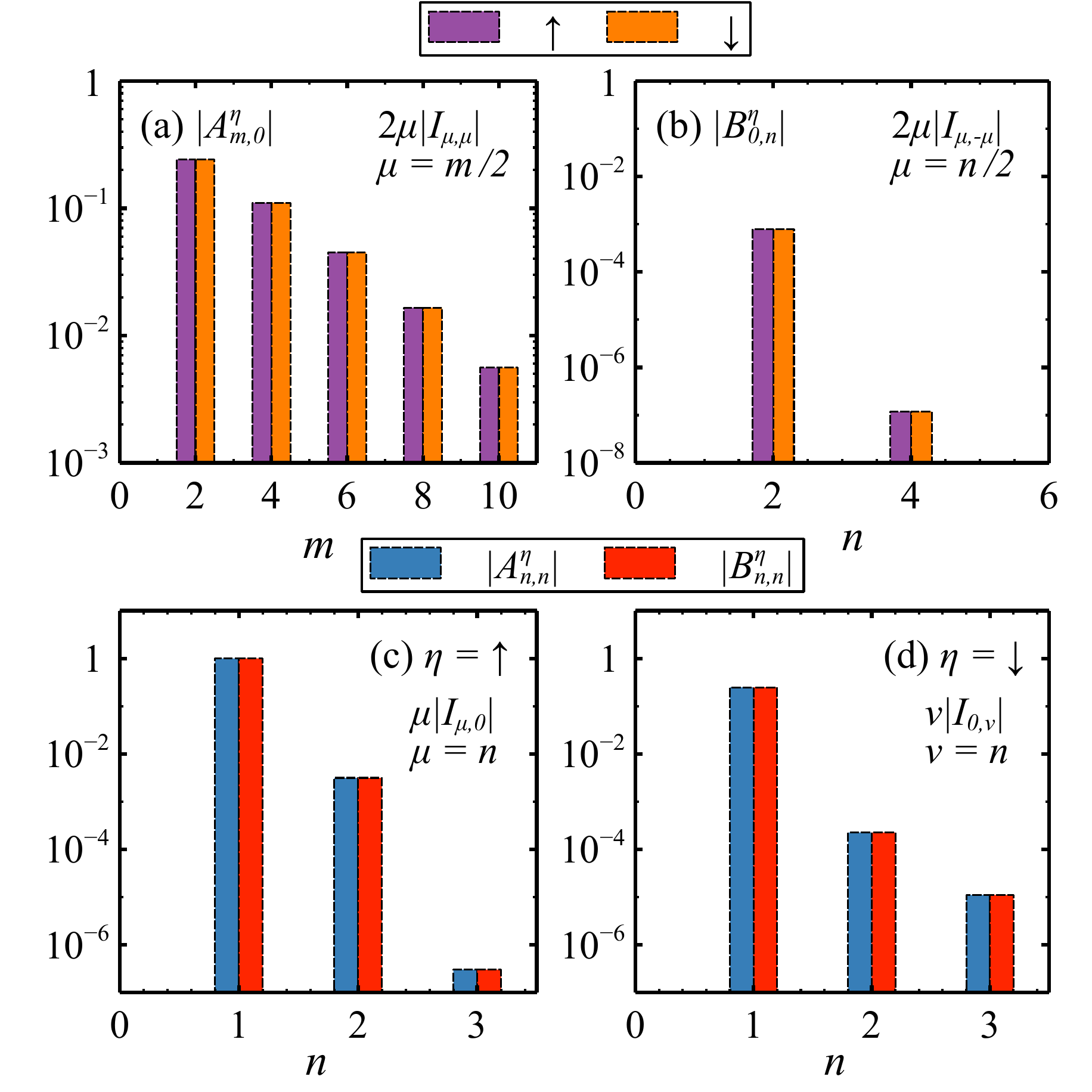}
    \caption{Fourier coefficients obtained from Eqs.~\eqref{eqn:general_Fourier_ansatz}-\eqref{eqn:Bmn}. Panel (a): The $A^\eta_{m,0}$ Fourier coefficients which determine the crossed-pair transmission processes, $I_{\mu,\mu}~(\mu=m/2)$ that turn our to be crucial for the JDE. Panel
    (b) The $B^\eta_{0,n}$ coefficients determining the processes of equal-spin pairs transmitted in the opposite directions, $I_{\mu,-\mu}~(\mu=n/2)$. In both panels, the purple/orange bars denote the $\uparrow\!/\!\downarrow$ transmission channel. Panels (c) and (d): The $A^\eta_{nn}$ (blue bars) and $B^\eta_{nn}$ (red bars) coefficients which determine the transmission of $n$ pairs in only one spin channel [panel (c) for $\uparrow$ and panel (d) for $\downarrow$]. In all panels, the coefficients are scaled with $|A^\uparrow_{1,1}|$. The analysis is performed for {the} parameters as in Fig.~\ref{fig:CPR_normal}}
    \label{fig:Foureir}
\end{figure}

Let us first consider the $A^\eta_{m,0}$ and $B^\eta_{0,n}$ coefficients, whose absolute values are shown in Figs.~\ref{fig:Foureir}(a) and~\ref{fig:Foureir}(b), respectively. Apparently, both coefficients coincide for spin-$\uparrow$ (purple bars) and spin-$\downarrow$ (orange bars) channels. We present the absolute values to make the following symmetry relations better visible:
\begin{equation}\label{eqn:A0}
    A^\uparrow_{m,0}=A^\downarrow_{m,0}\quad\text{and}\quad  B^\uparrow_{0,n}=-B^\downarrow_{0,n}.
\end{equation}
In addition, we note that only the even coefficients are nonvanishing, which corresponds to the so-called crossed-pair transmission processes. Namely, inserting Eq.~\eqref{eqn:A0} into Eq.~\eqref{eqn:I_mu}, we obtain $\mu I_{\mu,\mu}=\frac{1}{2}A^\eta_{2\mu,0}$, which corresponds to the coherent transport of $\mu$ equal-spin pairs in both spin bands in the same direction. In contrast, inserting Eq.~\eqref{eqn:A0} into Eq.~\eqref{eqn:I_nu} leads to $\nu I_{\nu,-\nu}=\frac{1}{2}B^\uparrow_{0,2\nu}$ corresponding to the coherent transport of $\nu$ equal-spin pairs in both spin bands in the opposite directions. Note that such processes do not contribute to the charge current. 

Let us now consider the $A^\eta_{n,n}$ and $B^\eta_{n,n}$ coefficients whose absolute values are shown in Fig.~\ref{fig:Foureir}(c) (spin-$\uparrow$ channel) and Fig.~\ref{fig:Foureir}(d) (spin-$\downarrow$ channel). Similarly as before, we present the absolute values corresponding to the following symmetry relations:
\begin{equation}\label{eqn:A_nn}
    A^\uparrow_{n,n}=-B^\uparrow_{n,n}\quad\text{and}\quad A^\downarrow_{n,n}=B^\downarrow_{n,n}.
\end{equation}
Comparing these equations with Eqs.~\eqref{eqn:I_mu} and~\eqref{eqn:I_nu} brings us to the conclusion that $A^\eta_{n,n}$ describes the coherent transport of $n$ equal-spin pairs in only one spin band, while zero in the other. Namely, $(-1)^\mu\mu I_{\mu,0}=A^\uparrow_{\mu,\mu}$ is the contribution from $\mu$ pairs transferred via spin-$\uparrow$ channel, and $(-1)^\nu\nu I_{0,\nu}=A^\downarrow_{\nu,\nu}$ is the contribution from $\nu$ pairs transferred via spin-$\downarrow$ channel.

\subsection{Harmonic approximation}
\begin{figure}[t!]
    \centering
    \includegraphics[width=0.75\linewidth]{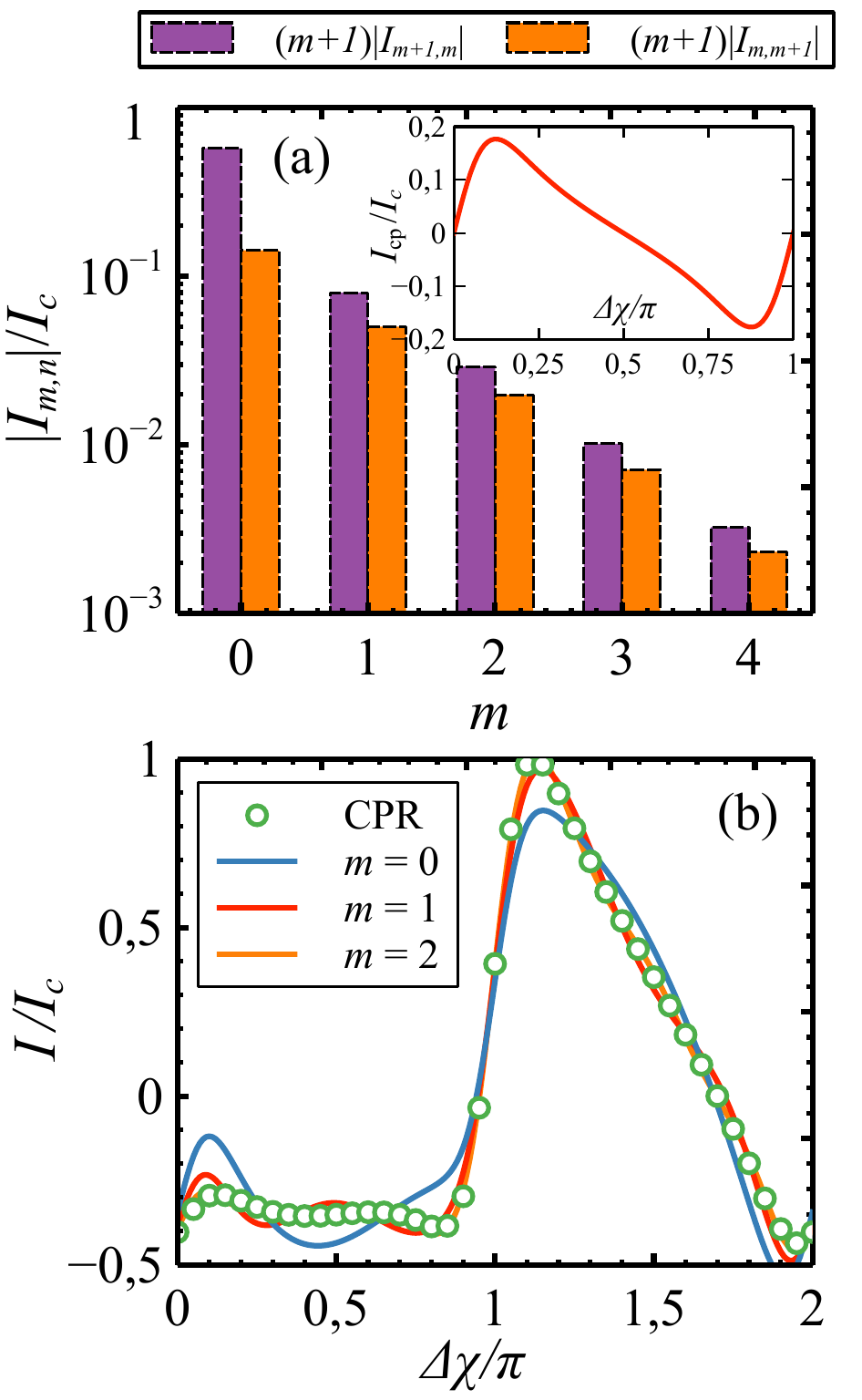}
    \caption{Panel (a): The $I_{m,m+1}$ (purple bars) and $I_{m+1,m}$ (orange bars) coefficients from the harmonic expansion of Eqs.~\eqref{eqn:I_up_approx} and~\eqref{eqn:I_down_approx}. The inset shows the crossed-pair current [see Eq.~\eqref{eqn:I_cp}]. Panel (b): The Josephson CPR (green circles) approximated by Eqs.~\eqref{eqn:I_up_approx} and~\eqref{eqn:I_down_approx} up to the first (blue line), the second (red line), and the third order (orange line). Parameters are the same as in Fig.~\ref{fig:Foureir}.}
    \label{fig:Approx}
\end{figure}
The above analysis can be utilized for describing the Josephson current-phase relation in terms of transferred pairs across the junction. As we show, it turns out that typically just certain processes significantly contribute to the CPR. Namely, in highly transmissive symmetric junctions, the main contributions to the supercurrent come from the crossed-pair transport (transport of an equal number of pairs in both spin channels) and processes involving the transport of $2m+1$ pairs, where one spin channel mediates $m$ and the other $m+1$ pairs. This situation means that we keep only the first moments in $\Delta\varphi^\prime$ in the harmonic expansion~\eqref{eqn:general_Fourier_ansatz}. Furthermore, noting that the $B_{0,2n}$ coefficients are small [see Fig.~\ref{fig:Foureir}(b)] and taking into account symmetries from Eq.~\eqref{eqn:A_nn}, we can write down 
\begingroup
\allowdisplaybreaks
\begin{align}
I_\eta \approx &A^\eta_{1,1}\sin(\Delta\chi-\zeta_\eta\Delta\varphi^\prime)+\frac{1}{2}\sum_{m\in 2\mathbb{N} } A^\eta_{m,0}\sin(m\Delta\chi) \nonumber\\
+&\frac{1}{2}\sum_{m\in2\mathbb{N}+1}(A^\eta_{m,1}+B^\eta_{m,1})\sin(m\Delta\chi+\Delta\varphi^\prime)+\nonumber\\
+&\frac{1}{2}\sum_{m\in2\mathbb{N}+1}(A^\eta_{m,1}-B^\eta_{m,1})\sin(m\Delta\chi-\Delta\varphi^\prime).\nonumber 
\end{align}
\endgroup
The first term corresponds to the transfer of one pair in only one spin channel, the second term describes the crossed-pair processes where an even number of pairs are transferred with equal contribution from both spin channels, and the third/fourth term corresponds to an odd number of pairs transmitted, such that spin-$\uparrow\!/\!\downarrow$ channel carries one pair more. Translated into the language of the number of transmitted pairs [see Eqs.~\eqref{eqn:I_mu} and~\eqref{eqn:I_nu}] the above relation reads
\begingroup
\allowdisplaybreaks
\begin{align}
    \label{eqn:I_up_approx}
    I_\uparrow &\approx I_\mathrm{cp}(2\Delta\chi) - \sum_{m=0}^\infty m I_{m,m+1}\sin[(2m+1)\Delta\chi+\Delta\varphi^\prime]\nonumber\\
    &-\sum_{m=0}^\infty(m+1) I_{m+1,m}\sin[(2m+1)\Delta\chi-\Delta\varphi^\prime],\\
    \label{eqn:I_down_approx}
    I_\downarrow &\approx I_\mathrm{cp}(2\Delta\chi) - \sum_{m=0}^\infty m I_{m+1,m}\sin[(2m+1)\Delta\chi-\Delta\varphi^\prime]\nonumber\\
    &-\sum_{m=0}^\infty (m+1) I_{m,m+1}\sin[(2m+1)\Delta\chi+\Delta\varphi^\prime],
\end{align}
\endgroup
where $I_\mathrm{cp}$ denotes the crossed-pair current, which is the same for both spin channels and contains only even harmonics
\begin{equation}\label{eqn:I_cp}
    I_{\rm cp}(2\Delta\chi)=\sum_m m I_{m,m}\sin(2m\Delta\chi).
\end{equation}
As we show immediately, keeping just a few terms in Eqs.~\eqref{eqn:I_up_approx} and~\eqref{eqn:I_down_approx} is sufficient for quantitative recover of the full CPR. Figure~\ref{fig:Approx}(a) shows the absolute value of the coefficients in the expansion for several $m$'s. It is supplemented by the inset showing the crossed-pair current $I_\mathrm{cp}$ as a function of the superconducting phase difference. We notice that the coefficients are roughly exponentially suppressed with increasing $m$. Panel (b) shows the validity of the approximation. Namely, we explicitly see that keeping only one coefficient in the expansion is sufficient for a qualitative description (blue line), while the third-order expansion yields a quantitative match (orange line). Note that in the crossed-pair current, it is necessary to include higher harmonics as well, which is the main difference from the diffusive limit~\cite{schulz2025_prl,schulz2025_prb}, where it is sufficient to keep only the lowest, $\sin(2\Delta\chi)$, contribution. This, however, does not alter the main conclusions.   

\section{Conclusions}\label{sec:Conclusions}
In summary, we have presented a theoretical study of the spin-resolved Josephson diode effect in strongly spin-polarized magnetic materials with inhomogeneous spin textures coupled to two BCS superconductors. The theory has been formulated in the framework of quantum Gor'kov and quasiclassical Eilenberger Green's function formalism, adapted to strongly spin-polarized materials in the adiabatic limit. The former allows neglecting the spin-mixed correlations (spin singlets and spin triplets with $s_z=0$), while the latter gives rise to the adiabatic spin geometric phase $\Delta\varphi_s$. This phase enters the Josephson CPR in a very similar manner to the superconducting phase difference, $\Delta\chi$, resulting in the anomalous and the Josephson diode effect. Our model predicts a significant JDE with a diode efficiency larger than $40\%$. To get a better insight, we have performed the harmonic analysis of the CPR. We have shown that the spin geometric phase plays an equivalent role as the quantum geometric phase, $\Delta\varphi$, that sets in as a consequence of the noncoplanar arrangement between the magnetization of the FM and the magnetic moments of the spin-active SC/FM interfaces. These two phases we have joined and termed the \textit{total spin geometric phase}. Finally, we have provided a physically appealing and intuitive picture of the effect in terms of the number of coherently transferred Cooper pairs across the junction. 
Based on our previous work on homogeneous ferromagnetic systems~\cite{schulz2025_prl,schulz2025_prb}, we expect the effect to be robust against impurities in the system. In addition, we do not expect the self-consistent treatment of the superconductors to alter the main conclusions qualitatively. Our theory can be tested in experimental setups involving conical magnetic materials such as Holmium. 
\acknowledgments
DN acknowledges support by the COST Action SUPERQUMAP, CA2114 and the University of Bordeaux and LOMA UMR-CNRS 5798 for hospitality during his visit. NLS and ME acknowledge funding by the Deutsche Forschungsgemeinschaft (DFG; German Research Foundation) under project number 530670387.

\appendix
\section{Boundary conditions for coherence functions}\label{Appendix:BC}

The boundary conditions used throughout are based on the quantum S-matrix approach developed in Refs.~\cite{eschrigScatteringProblemNonequilibrium2009,greinSpinDependentCooperPair2009,greinTheorySuperconductorferromagnetPointcontact2010}. In this approach, the central role is played by the renormalized Green's function defined in channel space and incorporating elementary processes that occur between two regions
\begin{align}
    \bm{\gamma}^\prime\,= \bm{S}\bm{\gamma}\bm{\tilde{S}},\\
    \bm{\tilde\gamma}^\prime\,= \bm{\tilde{S}}\bm{\tilde\gamma}\bm{S},
\end{align}
where $\bm{S}$ is the S-matrix and $\bm{\gamma}=\gamma_k\delta_{kk'}$ with $k$ denoting the scattering channel. In our particular case, we consider SC/FM interface with a strongly polarized ferromagnet. This means that in the SC has a single doubly spin degenerated channel  and the FM has two distinct channels, i.e.,
\begin{equation}
    \bm{\gamma} = \left(\begin{array}{ccc}
         \hat{\gamma}_1 & {} & {}\\
         & \gamma_2 &  \\
         & & \gamma_3
    \end{array}\right)
\end{equation}
where index $\eta=1$ corresponds to SC and the indices $\eta=2,3$ correspond to the $\uparrow$ and $\downarrow$ spin bands in the FM, respectively. In total, $\bm{\gamma}$ has a $4\times 4$ matrix structure. Similarly, the S-matrix has the following form [see Eq.~\eqref{eqn:S_matrix}]:
\begin{equation}\label{eqn:S_matrix_appendix}
    \bm{S} = \left(\begin{array}{ccc}
       \hat{R}_1 & \vv{T}_{12} & \vv{T}_{13} \\
        \vv{T}^T_{21} &  r_{22} & r_{23} \\
        \vv{T}^T_{31} &  r_{32} & r_{33}
    \end{array}\right),
\end{equation}
and the corresponding hole-like scattering matrix, $\bm{\tilde{S}}$, is obtained simply by $\tilde{S}_{ij}=S_{ij}^\ast$. As mentioned in the main text, the $\hat{R}$ block describes the reflection (normal and spin-flip) in the SC, $r_{ij}$ amplitudes correspond to reflections in the FM, and $\vv{T}_{ij}=(t_{ij},t_{ij}^\prime)$ denotes the transmission amplitudes.

Following the notation of Ref.~\cite{eschrigScatteringProblemNonequilibrium2009}, we denote the incoming amplitudes with small letters, $\gamma$ and $\tilde\gamma$, and the outgoing ones with the capital ones, $\Gamma$ and $\tilde\Gamma$. However, since $\gamma$ and $\tilde\gamma$ have the opposite group (Fermi) velocities, we should also distinguish between the incoming and outgoing directions denoted by $p,p^\prime, \dots$ and $k,k^\prime,\dots$, respectively. Considering the \textit{outgoing directions}, the quasiclassical GF is built from $\tilde\gamma$ and $\Gamma$, i.e.,
\begin{equation}
    \hat{g}_k=\hat{g}_k[\Gamma,\tilde\gamma].
\end{equation}
To obtain the outgoing $\Gamma$ amplitude, we follow Ref. \cite{eschrigScatteringProblemNonequilibrium2009} in introducing the following functions expressed in a compact matrix form in channel space:
\begin{eqnarray}
    \bm{\mathcal{F}} &&\,= \bm{\gamma}^\prime(\bm{1}-\bm{\tilde\gamma}\bm{\gamma}^\prime)^{-1}=\bm{\mathcal{G}}\bm{\gamma}^\prime,\\
    \bm{\mathcal{G}} &&\,= (\bm{1}-\bm{\gamma}^\prime\bm{\tilde\gamma})=\bm{1}+\bm{\mathcal{F}\bm{\tilde\gamma}}.
\end{eqnarray}
The quasiclassical amplitudes are $\mathcal{G}_k=\mathcal{G}_{kk}$ and $\mathcal{F}_k=\mathcal{F}_{kk}$ and the outgoing amplitude is calculated as
\begin{equation}\label{eqn:Gamma_out}
    \Gamma_k = [\mathcal{G}_k]^{-1}\mathcal{F}_k.
\end{equation}
A similar consideration can be performed for the \textit{incoming directions} where the quasiclassical GF is
\begin{equation}
    \hat{g}_p=\hat{g}_p[\gamma,\tilde\Gamma].
\end{equation}
The outgoing $\tilde\Gamma$ amplitude is calculated with the help of
\begin{eqnarray}
    \bm{\tilde\mathcal{F}} &&\,= \bm{\tilde\gamma}^\prime(\bm{1}-\bm{\gamma}\bm{\tilde\gamma}^\prime)^{-1}=\bm{\tilde\mathcal{G}}\bm{\gamma}^\prime,\\
    \bm{\tilde\mathcal{G}} &&\,= (\bm{1}-\bm{\tilde\gamma}^\prime\bm{\gamma})=\bm{1}+\bm{\tilde\mathcal{F}\bm{\gamma}},
\end{eqnarray}
as
\begin{equation}\label{eqn:GammaTilde_out}
    \tilde\Gamma_p = [\tilde{\mathcal{G}}_p]^{-1}\tilde{\mathcal{F}}_p.
\end{equation}
These boundary conditions are used for calculating the Josephson current in Sec.~\ref{sec:Results} in the main text. 

\section{Linearized boundary conditions in the tunneling limit}\label{Appendix:Linearized_BC}
Equations~\eqref{eqn:Gamma_out} and  \eqref{eqn:GammaTilde_out} can be rewritten as the Dyson-like equations, respectively~\cite{eschrigScatteringProblemNonequilibrium2009}:
\begin{align}
    \Gamma_{k\leftarrow k'} = \gamma^\prime_{kk'}+\sum_{k_1\neq k}\Gamma_{k\leftarrow k_1}\tilde{\gamma}_{k_1}\gamma^\prime_{k_1k'} \\
    \tilde{\Gamma}_{p\leftarrow p'} = \tilde{\gamma}^\prime_{pp'} + \sum_{p_1\neq p}\tilde{\Gamma}_{p\leftarrow p_1}\gamma_{p_1}\tilde{\gamma}^\prime_{p_1p'}
\end{align}
Following the notation in which the scattering channel of SC is labeled by $\eta=1$ and the two channels of FM by $\eta=2,3$ (corresponding to $\sigma=\uparrow\downarrow$ respectively), the outgoing amplitude into the spin-$\uparrow$ band of FM is given by
\begin{align}\label{eqn:Gamma2}
    \Gamma_2 = \gamma_{22}^\prime + \Gamma_{2\leftarrow 1}\, \tilde{\gamma}_1\, \gamma_{12}^\prime + \Gamma_{2\leftarrow 3}\, \tilde{\gamma}_3\, \gamma_{32}^\prime.
\end{align}
Here,
\begin{align}
    \Gamma_{2\leftarrow 1} &= \qty[\gamma_{21}^\prime + \gamma_{23}^\prime N_3 \tilde{\gamma}_3 \gamma_{31}^\prime] N_1 N_{13}, \\
    \Gamma_{2\leftarrow 3} &= \qty[\gamma_{23}^\prime + \Gamma_{2\leftarrow 1} \tilde{\gamma}_1 \gamma_{13}^\prime] N_3,
\end{align}
with
\begin{equation}
\begin{split}
    &N_i = \qty(1-\tilde{\gamma}_i\gamma_{ii}^\prime)^{-1}, \\
    &N_{ij} = \qty(1-\tilde{\gamma}_i \gamma_{ij}^\prime N_j \tilde{\gamma}_j \gamma_{ji}^\prime N_i)^{-1}.
\end{split}
\end{equation}
An analogous set of equations for the scattering into the spin-$\downarrow$ band can be obtained by replacing $2\leftrightarrow 3$. 

If we neglect the terms $\sim\mathcal{O}(t_{ij}^3)$, Eq.~\eqref{eqn:Gamma2} significantly simplifies adopting the form 
\begin{equation}
    \label{eqn:Gamma2_tunneling}
    \Gamma_2=\gamma^\prime_{22}+\gamma^\prime_{21}(1-\tilde{\hat\gamma}_1\gamma^\prime_{11})^{-1}\tilde{\hat\gamma}_1\gamma^\prime_{12}.
\end{equation}
Now we keep only terms linear in $\gamma_{22}$ and $\gamma_{33}$ and quadratic in $t_{ij}$ arriving at
\begin{equation}\label{eqn:Gamma2_tunneling_2}
    \Gamma_2= |r_{22}|^2\gamma_2 + r_{23}r_{32}^\ast\gamma_3 + A_2,
\end{equation}
where 
\begin{equation}
    A_2\!=\!\vv{T}_{21}^T\qty [\hat{\gamma}_1\!+\!\hat{\gamma}_1\hat{R}_1^\ast(1\!-\!\tilde{\hat\gamma}_1\hat{R}_1\hat{\gamma}_1\hat{R}_1^\ast)^{-1}\tilde{\hat{\gamma}}_1\hat{R}_1\hat{\gamma}_1]\vv{T}_{12}.
\end{equation}
The first two terms on the r.h.s. of Eq.~\eqref{eqn:Gamma2_tunneling_2} represent the spin-conserving and spin-flip contributions, respectively, whereas $A_2$ can be viewed as a source term due to the transmission from the SC. We now apply the above discussion to the case of two SC/FM interfaces labeled by L and R. Expressing the incoming amplitudes at one interface by the outgoing ones at the other, we arrive at \cite{greinSpinDependentCooperPair2009}
\begin{align}
    \Gamma_{2}^L = |r^L_{22}|^2\beta_2^\ast\Gamma_{2}^R + r_{23}^Lr_{32}^{L\ast}\beta_3^\ast\Gamma_{3}^R + A_{2}^L,\\
    \Gamma_{2}^R = |r^R_{22}|^2\beta_2\Gamma_{2}^L + r_{23}^Rr_{32}^{R\ast}\beta_3\Gamma_{3}^L + A_{2}^R.
\end{align}
A similar set of equations can be obtained for spin-$\downarrow$ band of FM by the replacement $2\leftrightarrow 3$. Grouping the equations with respect to the corresponding interface and expressing them in a matrix form, we arrive at Eqs.~\eqref{eqn:Gamma} of the main text.

\section{Time reversal and inversion symmetries of the conical state}\label{Appendix:Symmetry}
As stated in the main text, the conical spin texture considered throughout has the form
\begin{equation}
     \vec{J}(z) = [\sin{\alpha}\cos{\phi(z)},\sin{\alpha}\sin{\phi(z)},\cos{\alpha}],
\end{equation}
giving rise to the following adiabatic spin gauge field
\begin{equation}
    \vec{Z}=-\frac{1}{4}(1-\cos{\alpha})\partial_z\phi(z)\vec{e}_z.
\end{equation}
Let us now consider how this field transforms under the time reversal and inversion operations. 

The time-reversal operator is antiunitary acting as
\begin{equation}
    \hat{\mathcal{T}}t\mapsto-t,
\end{equation}
and has a two-dimensional representation 
\begin{equation}
    \hat{\mathcal{T}}=-i\hat{\sigma}_2K,
\end{equation}
where $K$ is the complex conjugation operator. It does not leave the spin invariant, and the Pauli matrices transform as
\begin{equation}
    \hat{\mathcal{T}}\hat{\sigma}_i\hat{\mathcal{T}}^{-1}=-\hat{\sigma}_i.
\end{equation}
Consequently, the exchange interaction term transforms as
\begin{equation}
    \hat{\mathcal{T}}(\vec{J}\cdot\bm{\hat{\sigma}})\hat{\mathcal{T}}^{-1} = -\vec{J}\cdot\bm{\hat{\sigma}}.
\end{equation}
Finally, this time reversal transformation leads to the following transformation of the spin-gauge field:
\begin{eqnarray}
    \vec{Z}\,&&=-\frac{1}{4}(1-\cos\alpha)\partial_z\phi\vec{e}_z \to \frac{1}{4}(1-\cos\alpha)\partial_z\phi\vec{e}_z=\nonumber \\
    &&=-\vec{Z}\implies \delta\varphi_s(\alpha,z)\to-\delta\varphi_s(\alpha,z).
\end{eqnarray}
By taking the linear modulation of the helix, $\phi(z)=qz$, the above transformation can be effectively seen as $q\to-q$. 

A similar consideration can be performed for the inversion (parity) symmetry defined as
\begin{equation}
    \hat{\Pi}\left(\begin{array}{ccc}
         x  \\
         y \\
         z
    \end{array}\right)\mapsto
    \left(\begin{array}{ccc}
         -x  \\
         -y \\
         -z
    \end{array}\right).
\end{equation}
Correspondingly, the spherical coordinates transform as
\begin{equation}
    \alpha \mapsto \pi-\alpha\quad\text{and}\quad\varphi\mapsto\varphi+\pi.
\end{equation}
Casting this transformation into the spin-exchange Hamiltonian leads to the following transformation of the exchange interaction:
\begin{equation}
    \hat{\Pi}(\vec{J}\cdot\bm{\hat{\sigma}})\hat{\Pi}^{-1}=-\vec{J}\cdot\bm{\hat{\sigma}},
\end{equation}
and, consequently, the unitary transformation reads
\begin{equation}
    \hat{U}(z)\mapsto\exp[i\frac{\pi-\alpha}{2}\vec{n}(-z)\cdot\bm{\hat{\sigma}}].
\end{equation}
Finally, the adiabatic spin gauge field adopts the form
\begin{align}
    \vec{Z}=-\frac{1}{4}(1-\cos\alpha)\partial_z\phi\vec{e}_z\to-\frac{1}{4}(1+\cos\alpha)\partial_z\phi\vec{e}_z
\end{align}
Note that the spin gauge phase remains invariant for $\alpha=\pi/2$ (helical state with zero tilt). In other words, in this case, the inversion symmetry is not broken which is obvious since we do not have an out-of-plane magnetization component. 

\section{Fourier coefficients}\label{Appendix:Fourier}
The spin-resolved currents transform under time reversal as
\begin{equation}\label{eqn:I_eta_time_reversal}
   I_\eta(\Delta\chi,\Delta\varphi, \Delta \varphi_s) = - I_\eta(-\Delta\chi,-\Delta\varphi,-\Delta\varphi_s), 
\end{equation}
leading to the Fourier expansion given in the following general form,
(we omit $\eta$ for compactness):
\begin{align}\label{eqn:I_Fourier_general}
    I(&\Delta\chi,\Delta\varphi,\Delta\varphi_s)=\\
    &=\sum_{m,n=0}^\infty \big[A_{m,n}^\prime (\Delta\varphi_s)\sin(m\Delta\chi)\cos(n\Delta\varphi)+\nonumber\\
    &\qquad\quad+B_{m,n}^\prime (\Delta\varphi_s)\cos(m\Delta\chi)\sin(n\Delta\varphi)+\nonumber\\
    &\qquad\quad+C_{m,n}^\prime (\Delta\varphi_s)\sin(m\Delta\chi)\sin(n\Delta\varphi)+\nonumber\\
    &\qquad\quad+D_{m,n}^\prime (\Delta\varphi_s)\cos(m\Delta\chi)\cos(n\Delta\varphi)\big].\nonumber
\end{align}
Note that Eq.~\eqref{eqn:I_eta_time_reversal} requires $A_{m,n}^\prime $ and $B_{m,n}^\prime $ to be even in $\Delta\varphi_s$ and $C_{m,n}^\prime $ and $D_{m,n}^\prime $ to be odd in $\Delta\varphi_s$.

To obtain Eq.~\eqref{eqn:general_Fourier_ansatz} of the main text, we shift $\Delta\varphi$ by $\Delta\varphi_s$, i.e., $\Delta\varphi\to\Delta\varphi-\Delta\varphi_s$ arriving at
\begin{widetext}
\begin{equation}\label{eqn:I_abcd}
\begin{split}
I (\Delta\chi,\Delta\varphi-\Delta\varphi_s,\Delta\varphi_s) = \sum_{m,n=0}^\infty\big\{&[A_{m,n}^\prime (\Delta\varphi_s)\cos(n\Delta\varphi_s)-C_{m,n}^\prime (\Delta\varphi_s)\sin(n\Delta\varphi_s)]\sin(m\Delta\chi)\cos(n\Delta\varphi)+\\
    +&[B_{m,n}^\prime (\Delta\varphi_s)\cos(n\Delta\varphi_s) + D_{m,n}^\prime (\Delta\varphi_s)\sin(n\Delta\varphi_s)]\cos(m\Delta\chi)\sin(n\Delta\varphi)+\\
    +&[C_{m,n}^\prime (\Delta\varphi_s)\cos(n\Delta\varphi_s)+A_{m,n}^\prime (\Delta\varphi_s)\sin(n\Delta\varphi_s)]\sin(m\Delta\chi)\sin(n\Delta\varphi)+\\
    +&[D_{m,n}^\prime (\Delta\varphi_s)\cos(n\Delta\varphi_s)-B_{m,n}^\prime (\Delta\varphi_s)\sin(n\Delta\varphi_s)]\cos(m\Delta\chi)\cos(n\Delta\varphi)\big\}.
\end{split}
\end{equation}
\end{widetext},
We find numerically that the last two  terms in the sum in Eq.~\eqref{eqn:I_abcd} vanish identically, leading to
the following identities:
\begingroup
\allowdisplaybreaks
\begin{align}
    0&\!\equiv\!C_{m,n}^\prime (\Delta\varphi_s)\!\cos(n\Delta\varphi_s)\!+\!A_{m,n}^\prime (\Delta\varphi_s)\!\sin(n\Delta\varphi_s),\!\\
     0&\!\equiv\!D_{m,n}^\prime (\Delta\varphi_s)\!\cos(n\Delta\varphi_s)\!-\!B_{m,n}^\prime\! (\Delta\varphi_s)\!\sin(n\Delta\varphi_s).\!
\end{align}
\endgroup
Furthermore, we find that the coefficients of the first two terms in Eq.~\eqref{eqn:I_abcd},
\begingroup
\allowdisplaybreaks
\begin{align}
    A_{m,n}=A_{m,n}^\prime (\Delta\varphi_s)\cos(n\Delta\varphi_s)-C_{m,n}^\prime (\Delta\varphi_s)\sin(n\Delta\varphi_s),\nonumber\\
    B_{m,n} = B_{m,n}^\prime (\Delta\varphi_s)\cos(n\Delta\varphi_s) + D_{m,n}^\prime (\Delta\varphi_s)\sin(n\Delta\varphi_s),
\end{align}
\endgroup
are independent of the variable $\Delta\varphi_s $. Expressing all coefficients by the $A_{m,n}$ and $B_{m,n}$,
\begingroup
\allowdisplaybreaks
\begin{align}
    &A_{m,n}^\prime (\Delta\varphi_s)=\quad \!A_{m,n} \cos(n\Delta\varphi_s),\\&B_{m,n}^\prime (\Delta\varphi_s)= \quad \! B_{m,n}\cos(n\Delta\varphi_s),\\
    &C_{m,n}^\prime (\Delta\varphi_s)=-A_{m,n} \sin(n\Delta\varphi_s),\\
     &D_{m,n}^\prime (\Delta\varphi_s)=\quad \! B_{m,n}\sin(n\Delta\varphi_s).
\end{align}
\endgroup
and introducing these expressions into
 Eq.~\eqref{eqn:I_Fourier_general}, we obtain
\begin{widetext}
\begin{align}
I (\Delta\chi,\Delta\varphi,\Delta\varphi_s) = \sum_{m,n=0}^\infty \left\{A_{m,n} \sin(m\Delta\chi) \cos[n(\Delta\varphi+\Delta\varphi_s)]+
B_{m,n}\cos(m\Delta\chi)\sin[n(\Delta\varphi+\Delta\varphi_s)] \right\}.
\end{align}
Introducing $\Delta\varphi'=\Delta\varphi+\Delta\varphi_s$, we find that $I$ in fact only depends on $\Delta \chi$ and $\Delta \varphi'$, $I (\Delta\chi,\Delta\varphi,\Delta \varphi_s)=
I (\Delta\chi,\Delta\varphi',0)$, such that 
\begin{align}
I (\Delta\chi,\Delta\varphi',0)\equiv I (\Delta\chi,\Delta\varphi')= \sum_{m,n=0}^\infty \left[A_{m,n} \sin(m\Delta\chi)\cos(n\Delta\varphi')+
B_{m,n}\cos(m\Delta\chi)\sin(n\Delta\varphi') \right].
\end{align}
\end{widetext}
This yields Eq.~\eqref{eqn:general_Fourier_ansatz} in the main text.

\end{document}